\documentclass[twocolumn]{aastex63}
\usepackage{amssymb,amsmath}
\usepackage{graphicx}
\usepackage{xcolor,natbib}
\usepackage{multirow}
\usepackage[T1]{fontenc}
\usepackage{ae,aecompl}
\usepackage{newtxtext, newtxmath}
\usepackage{float}
\usepackage{subfigure}
\usepackage{longtable}
\usepackage{enumitem}
\usepackage{longtable,listings}
\usepackage{longtable}
\usepackage{booktabs}
\usepackage{subfigure}
\usepackage{multirow}
\usepackage[flushleft]{threeparttable}
\usepackage{parcolumns}
\def\oiii{[O~{\sc iii}]}

\def\kms{${\rm km\,s}^{-1}$}

\submitjournal{ApJ}

\shorttitle{[O~{\sc iii}] line in type 1 AGNs}
\shortauthors{Zheng et al.}

\begin{document}
	
\title{Spectroscopic study of the broad component of [O~{\sc iii}]$\lambda5007$ profile in type 1 AGNs}

\correspondingauthor{Xueguang Zhang; Qirong Yuan}
\email{xgzhang@gxu.edu.cn; yuanqirong@njnu.edu.cn}

\author{Qi Zheng}
\affiliation{School of Physics and Technology, Nanjing Normal University, No. 1,	Wenyuan Road, Nanjing, 210023, P. R. China}

\author{Yansong Ma}
\affiliation{School of Physics and Technology, Nanjing Normal University, No. 1,	Wenyuan Road, Nanjing, 210023, P. R. China}

\author{Xueguang Zhang$^{*}$}
\affiliation{Guangxi Key Laboratory for Relativistic Astrophysics, School of Physical Science and Technology,
	GuangXi University, No. 100, Daxue Road, Nanning, 530004, P. R. China}

\author{Qirong Yuan$^{*}$}
\affiliation{School of Physics and Technology, Nanjing Normal University, No. 1,	Wenyuan Road, Nanjing, 210023, P. R. China}
\affiliation{University of Chinese Academy of Sciences, Nanjing 211135, P. R. China}

\begin{abstract}
The spectra of type 1 active galactic nuclei (AGNs) often exhibit broad component in \oiii$\lambda$5007, which are typically blue-shifted and associated with strong outflows.
We systematically analyze the \oiii~emission-line properties of type 1 AGNs with broad components to investigate how these kinematic features relate to the physical properties of the central engine.
From a parent sample of 11,557 QSOs at $z<0.3$ in Data Release 16 of the Sloan Digital Sky Survey, we select 2,290 type 1 AGNs exhibiting broad components in \oiii. 
Previous studies have reported a strong correlation between the blue emission, defined as the full extent of the broad component on the blue side, and black hole mass when the latter is estimated from the $M_{\rm BH}$–$\sigma_{\ast}$ relation using the line width $\sigma$ of the \oiii~core component as a surrogate for $\sigma_{\ast}$.
By the same way, the black hole mass also shows a strong correlation with the blue emission parameter in our sample.
However, this correlation becomes negligible when virial black hole masses are adopted. Besides, the velocity shifts between the broad and core components of \oiii~show a weak correlation with the Eddington ratio. It is consistent with the expectation that higher accretion rates enhance radiative pressure, thereby driving faster or more prominent outflows. 
In future work, we will compare \oiii~broad component properties between typical type 1 AGNs and those with double-peaked \oiii~to probe differences in narrow-line region kinematics and the impact of outflows or dual AGNs.
\end{abstract}

\section{Introduction}
Active galactic nuclei (AGNs) are among the most luminous and energetic phenomena in the universe, powered by the accretion of matter onto supermassive black holes (SMBHs) residing at the centers of galaxies \citep{Co06,Ma20,Ri23}. 
It is widely regarded that AGNs represent a crucial evolutionary phase in galaxy evolution during which the central black hole grows through accretion. 
In particular, AGN feedback is considered a key mechanism for establishing the correlations between the black hole and galaxy \citep{Fa12}.
By quenching both star formation and accretion onto the black hole, AGN feedback is thought to self-regulate BH growth and influence galaxy evolution \citep{Di05,Ho06,Ji25}.
A practical way to explore the observational manifestations of AGN feedback is to analyze the emission-line regions around the central black hole. These include the broad-line region (BLR) and the narrow-line region (NLR) \citep{An93,Ur95}, both of which are photoionized by the central engine.
Among these, the forbidden [O~{\sc iii}]$\lambda5007$ (hereafter \oiii) is especially significant due to its frequently observed asymmetry.
These emission lines generally exhibit two distinct components: a narrow core component that aligns with the systemic redshift of the host galaxy, and a broader component with higher line width, which is often significantly blue-shifted \citep{Cr10,Se21,Ko22}. This blue-shifted component is commonly interpreted as evidence of outflows \citep{Ch97,Gr05,Ne11}.

The \oiii~emission line, arising from a forbidden transition, can only originate in low-density gas. Therefore, any broadening or shifting observed in this line typically indicates significant velocity gradients or bulk motions within the NLR of AGNs \citep{Mu13}.
Unlike the dense BLR on sub-parsec scales, the NLR can extend over kiloparsecs \citep{De22}, making it a useful probe of this kinematics on galactic scales.

A substantial body of research has focused on the broad component of \oiii, encompassing both case studies of individual objects \citep{Cr07,Oh13} and statistical analyses based on large spectroscopic samples \citep{Mu13,Di18,Ko22,Ma23}.
Several studies have reported a correlation between the outflows traced by \oiii~and AGN luminosities in the optical \citep{Sc18}, infrared \citep{Za14,Za16,Su17}, and X-ray \citep{Pe17}, supporting that powerful outflows can be generated by the radiation pressure coming from the accretion disk \citep{Po00}.
In certain cases, however, outflows may be launched by radio jets \citep{Ro10,Mu13,Pe23}, or alternatively, the observed radio emission could result from shocks produced by radiatively driven winds \citep{Za14}.
Outflow velocities may also be linked to other AGN properties, such as black hole mass \citep{Ru17,Te20,Sc21}, accretion rate \citep{Gr05,Wo16}, and star formation \citep{Wo20,Ki22,Mu22}, although the evidence for such correlations remains weak or inconclusive in some studies \citep{Zh11,Pe14}.

In this study, we focus on the asymmetry of the \oiii~emission line in type 1 AGNs. 
Previous work suggested that type 1 AGNs may exhibit more prominent broad components than type 2 systems in \oiii, possibly due to reduced obscuration and projection effects along the line of sight \citep{Mu13}. 
After accounting for the projection effect, the fraction of outflows in type 2 AGNs is comparable to that reported for type 1 AGNs in earlier studies \citep{Ba14}.
More recently, \citet{To24} reported that, at fixed bolometric luminosity, type 2 AGNs may even host faster ionized outflows than their type 1 counterparts. 
Given these complexities and the potential for obscuration to bias kinematic measurements, we focus our analysis on type 1 AGNs.
Recent work by \citet{Sc18} revealed a correlation between the degree of \oiii~asymmetry and black hole mass in a sample of 28 narrow-line Seyfert 1 galaxies with blue-shifted broad components. A similar trend was later identified by \citet{Sc21} in a sample of 45 nearby southern broad-line Seyfert 1 galaxies. While these studies provide valuable insights into the kinematics of ionized gas outflows, their conclusions are limited by small sample sizes. Consequently, it remains challenging to assess whether such correlations hold across the broader population of type 1 AGNs.

Building on the work of \citet{Sc18,Sc21}, we aim to extend the investigation by adopting a complementary approach—examining the gas kinematics in a much larger sample of type 1 AGNs, encompassing both narrow-line Seyfert 1 and broad-line Seyfert 1 galaxies, from the Data Release 16 of the Sloan Digital Sky Survey (SDSS DR16) \citep{Ah20}. In particular, detailed studies of nearby Seyfert galaxies are crucial for deepening our understanding of AGN feedback and kinematics at higher redshifts, as these systems are expected to share similar physical properties across cosmic time.

In this manuscript, the selection procedure for type 1 AGNs with broad components in \oiii is described in Section 2, the basic physical parameters of this type 1 AGN sample are calculated in Section 3, the discussions about the results are shown in Section 4, 
and the summary and conclusions are provided in Section 5.
We have adopted the cosmological parameters of $H_{0}=70 {\rm km/s/Mpc}$, $\Omega_{\Lambda}=0.7$ and $\Omega_{\rm m}=0.3$.
\section{Sample Selection}
\begin{figure*}
\centering\includegraphics[width=5.8cm,height=3.6cm]{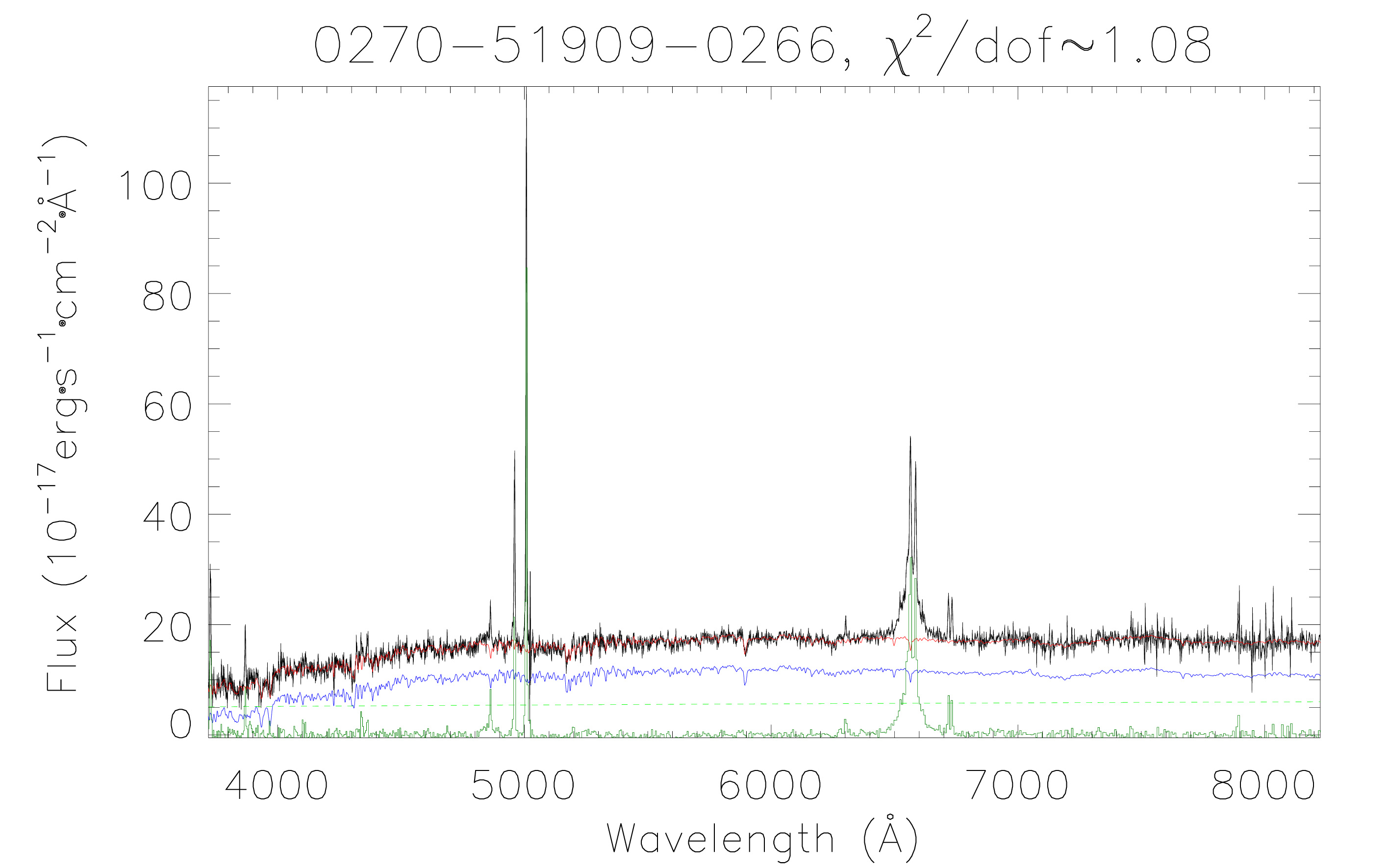}
\centering\includegraphics[width=5.8cm,height=3.6cm]{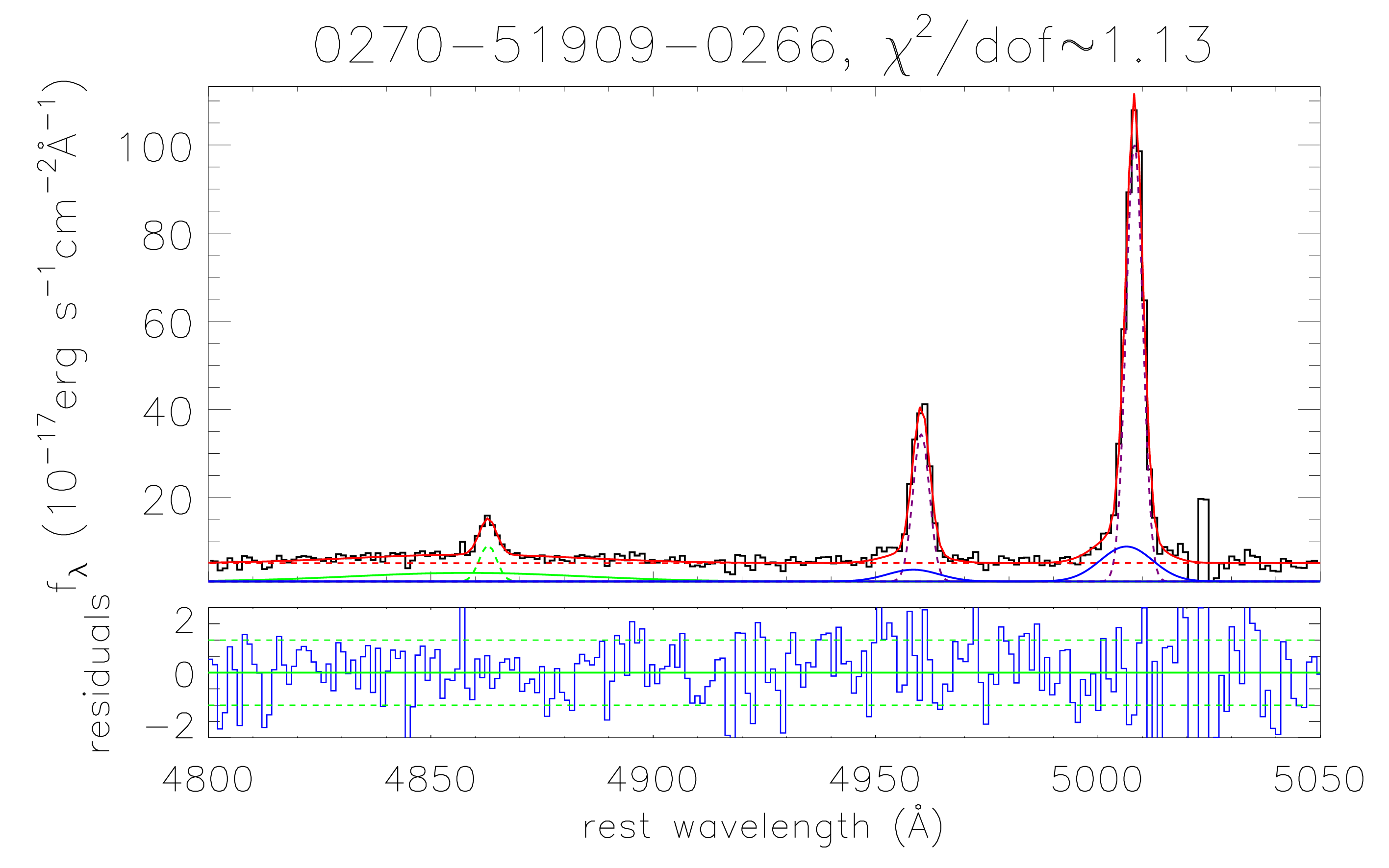}
\centering\includegraphics[width=5.8cm,height=3.6cm]{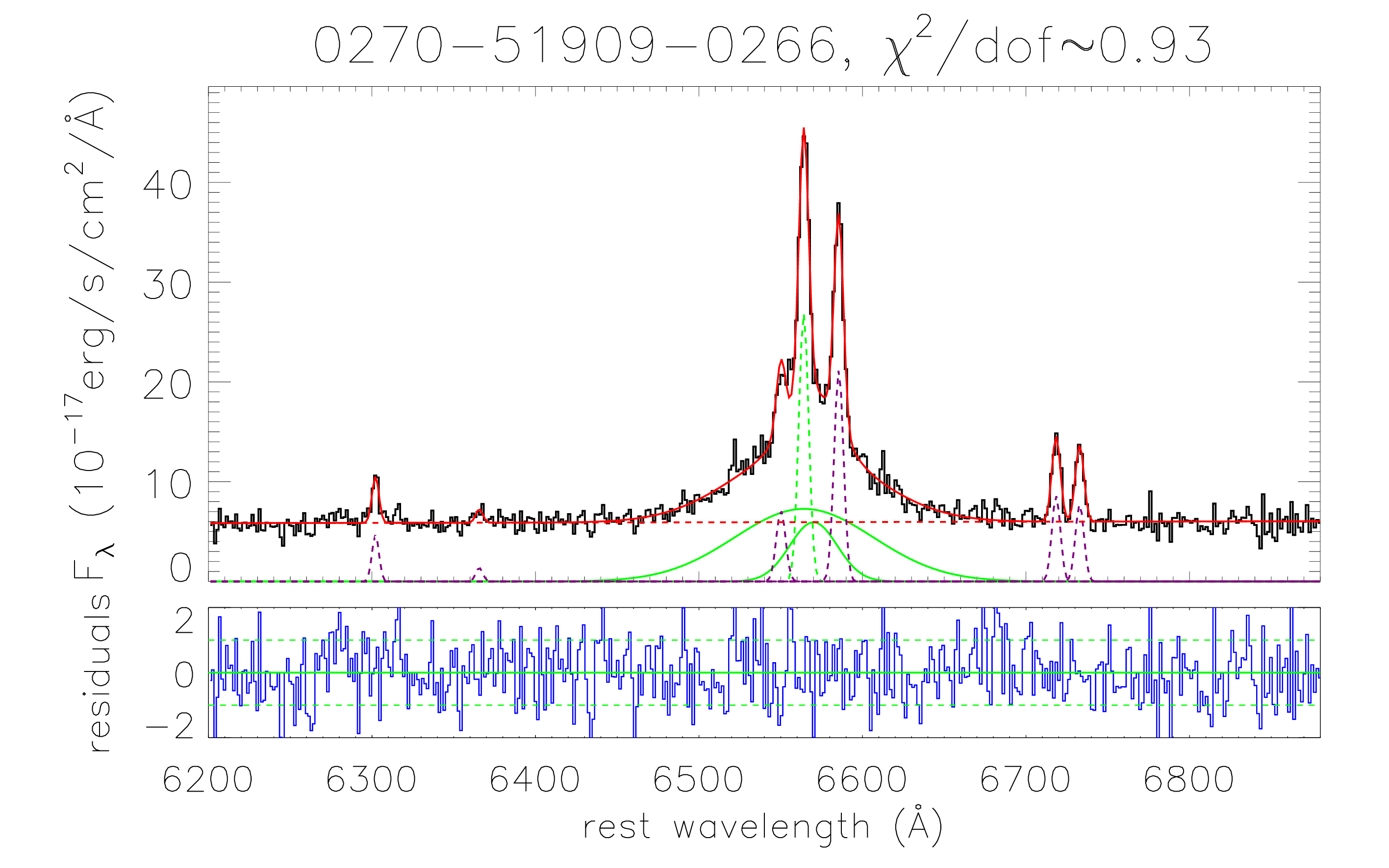}
\centering\includegraphics[width=5.8cm,height=3.6cm]{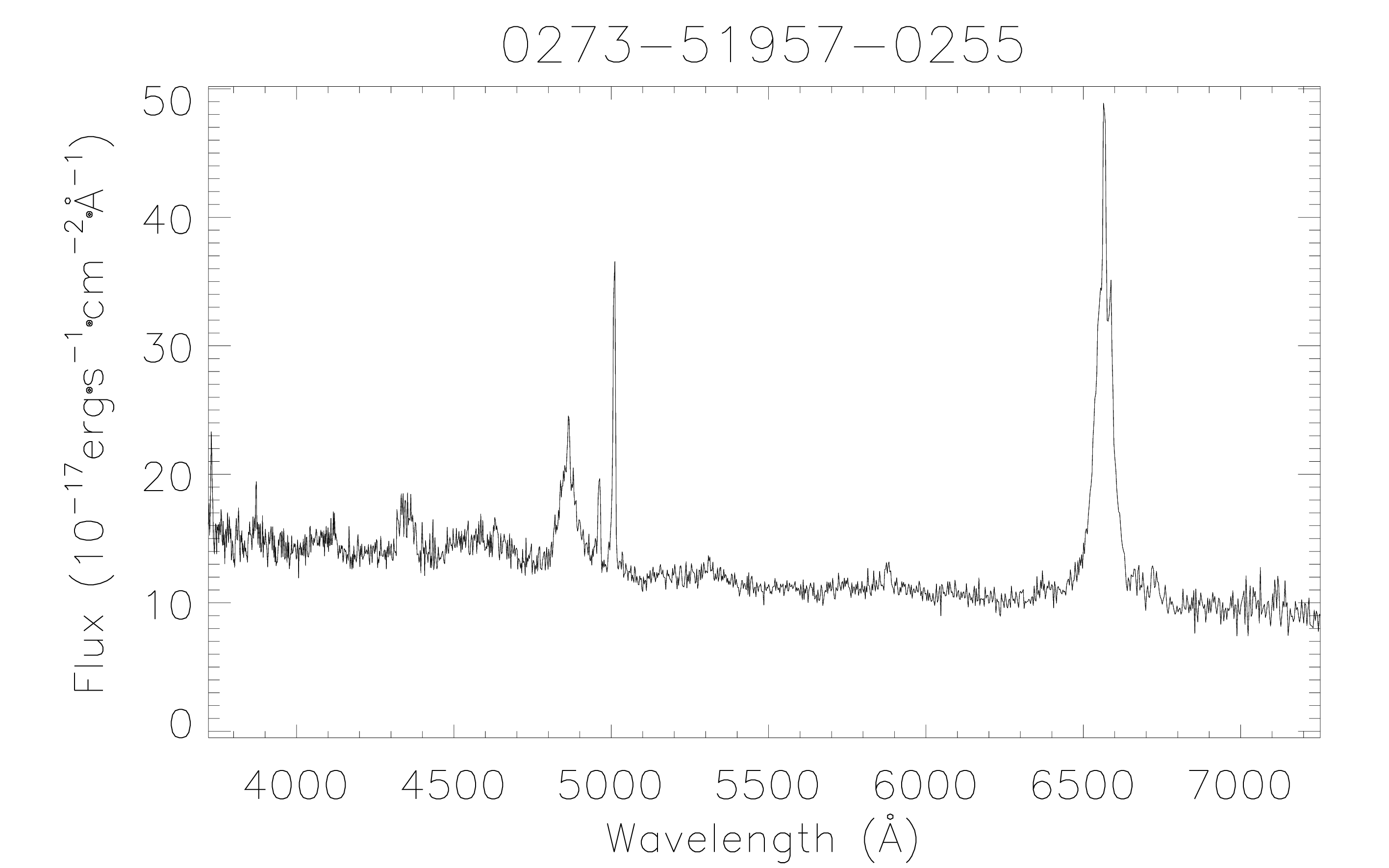}
\centering\includegraphics[width=5.8cm,height=3.6cm]{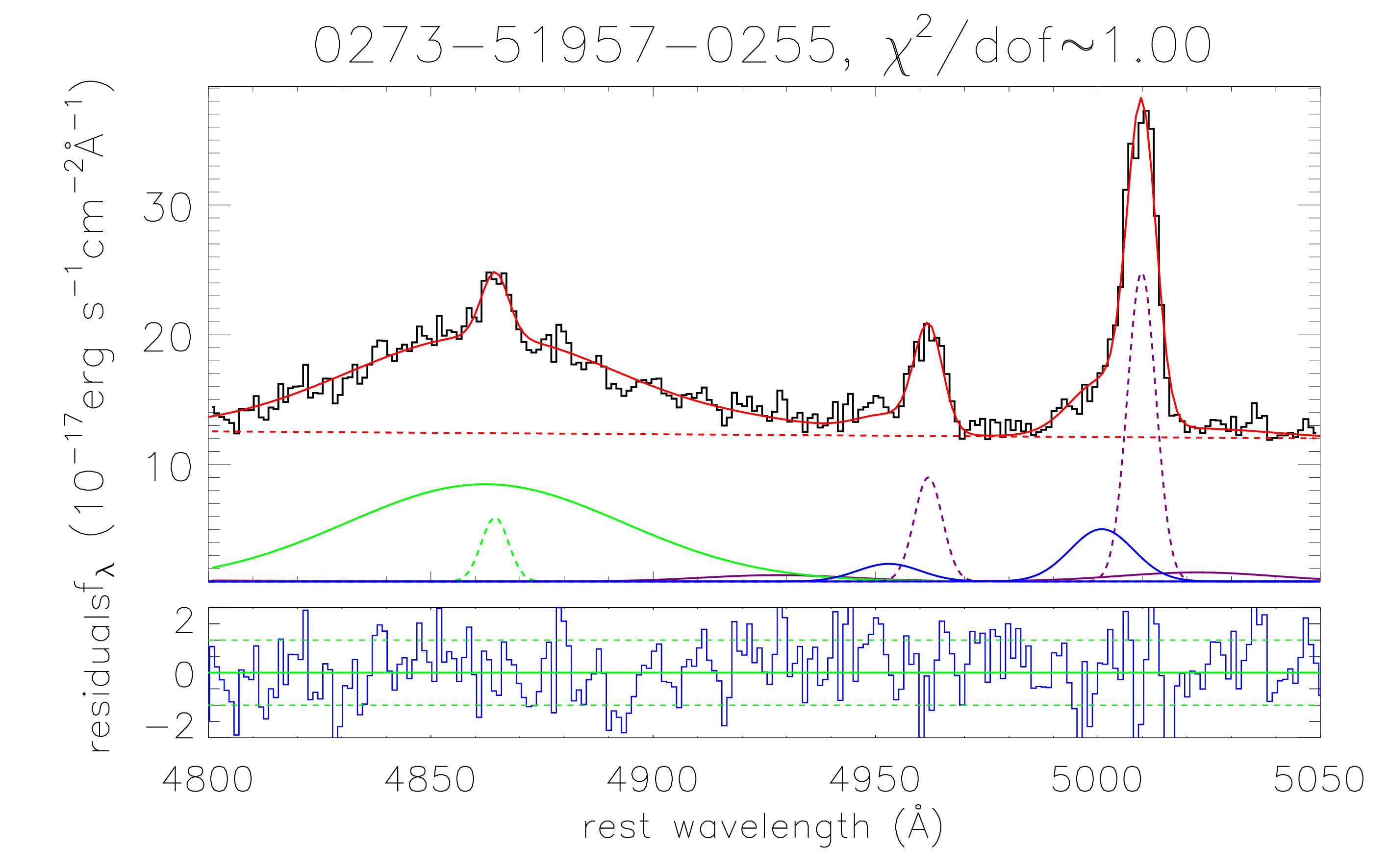}
\centering\includegraphics[width=5.8cm,height=3.6cm]{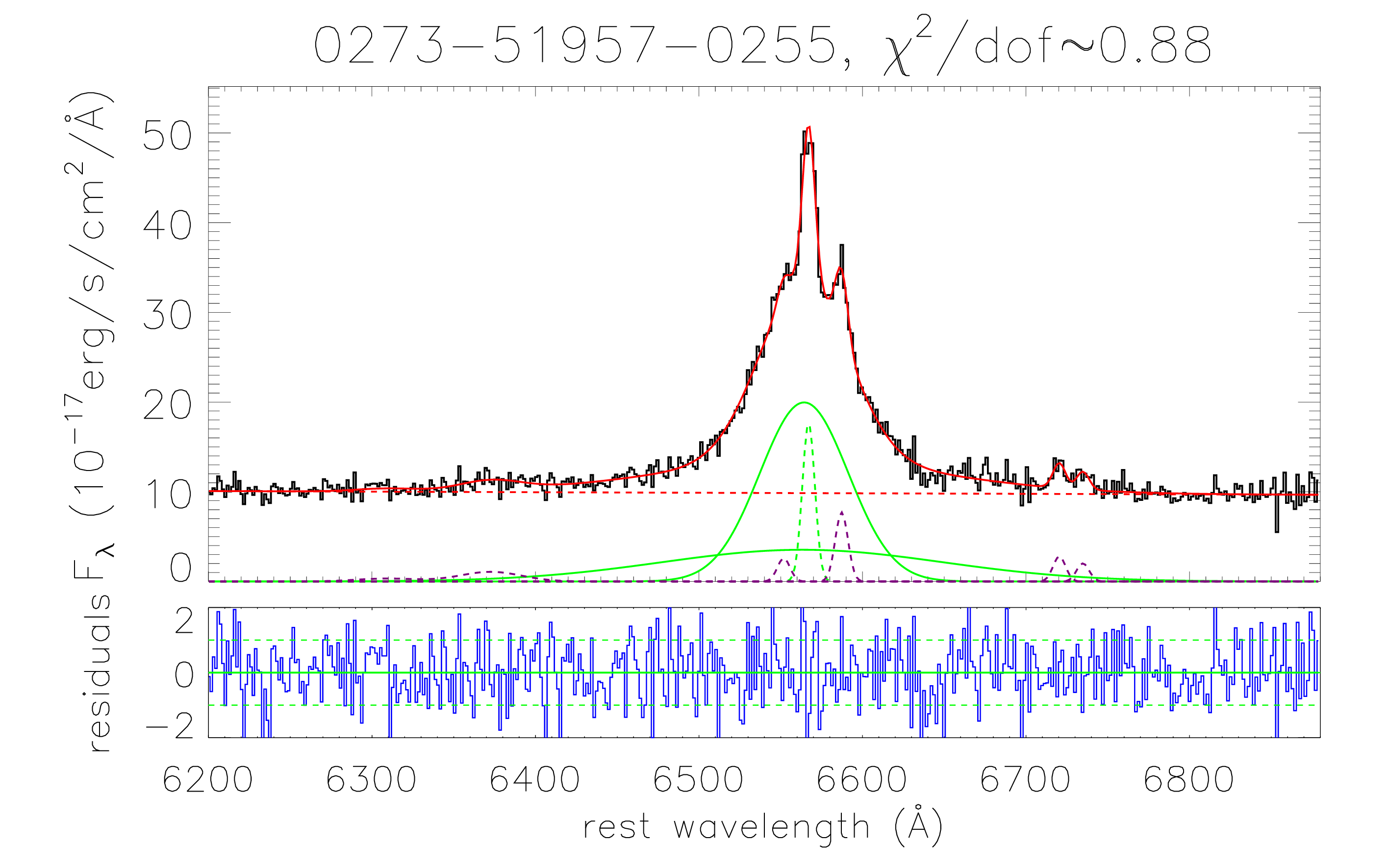}
\caption{Two examples of spectra and corresponding best fitting results in our sample. In each panel, the information of Plate-Mjd-Fiberid is shown in the title, the solid red line ($f_{best}$) shows the best fitting results with $\chi^2/\rm dof$ (if present) shown in the title.
In the left panels, the solid black lines show observed spectra.	
In the top-left panel, the solid blue line shows starlight, the dashed green line represents the AGN continuum emission, and the solid dark green line represents the line spectrum calculated by the SDSS spectrum minus the best fitting results. 
In the top of middle and right panels, the solid black lines ($f_{obs}$) show the line spectra after subtracting starlight (if present), the dashed red lines represent continuum emissions, the solid green lines represent broad H$\beta$ and H$\alpha$ emission lines, the dashed green lines represent narrow H$\beta$ and H$\alpha$ emission lines, the dashed purple lines represent core components of \oiii, [O~{\sc i}], [N~{\sc ii}] and [S~{\sc ii}] doublets, the solid blue lines represent the broad component of \oiii. 
In the bottom of middle and right panels, the solid blue lines represent the normalized residuals calculated by $(f_{obs}-f_{best})/err$, where $err$ is the uncertainty of SDSS spectrum, the horizontal solid and dashed green lines show residuals=0, ±1, respectively.
}
\label{fig0}
\end{figure*}

Same as down in \citet{Zh25}, we select 11,557 QSOs with $z<0.3$ from SDSS DR16 \citep{Ah20} as the parent sample using the SQL (Structured Query Language) query.
Then, to account for significant host galaxy contamination in the SDSS spectra of parent sample, we employ the established simple stellar population (SSP) method following \citep{Br93,Br03,Ka03,Ci05,ca17}.
Our implementation combines 39 broadened, strengthened and shifted SSP templates spanning 13 ages (5 Myr-12 Gyr) and 3 metallicities (Z=0.008, 0.05, 0.02), with a power-law component to model the AGN continuum.
More details about the same method can be found in \citet{zxg14,zxg21,Zh25}.
Then the observed spectrum with apparent host galaxy contribution is modeled by SSP templates plus a power-law component through the Levenberg-Marquardt least-squares minimization technique (the MPFIT package; \citealp{Ma09}).
In the left panel of Figure \ref{fig0}, we show one object with obvious host galaxy contributions and one object with no host galaxy contributions in our final sample.

Following possible host galaxy subtraction, we model the emission lines around H$\beta$ (rest wavelength from 4400 to 5600 \AA)
and H$\alpha$ (rest wavelength from 6200 to 6880 \AA).
The profile of H$\beta$ (H$\alpha$) consists of one narrow ($\sigma$ < 600 \kms) and two broad Gaussian functions ($\sigma$ > 600 \kms). 
[O{\sc i}], [S{\sc ii}] and [N{\sc ii}] doublets are each modeled with two Gaussian components sharing the same redshift and line width in velocity space, with the [N{\sc ii}] flux ratio fixed at its theoretical value of 3.
The  [O~{\sc iii}]$\lambda5007\AA$ line is modeled using two Gaussian components, representing the core and the broad components. Its [O~{\sc iii}]$\lambda4959\AA$ counterpart is constrained to share the same central wavelength and line width, and its flux is fixed at one-third of the  [O~{\sc iii}]$\lambda5007\AA$ flux, consistent with the theoretical ratio.
He~{\sc ii} line is represented by a single broad Gaussian component, while the optical Fe~{\sc ii} component is described using the optical Fe~{\sc ii} emission templates in the four groups discussed in \citet{kp10}.
The possible continuum emission beneath the emission lines around H$\beta$ (H$\alpha$) is modeled with a power-law component.
When fitting the H$\alpha$ (H$\beta$) profile, no additional broad component specifically associated with outflowing gas is included. 
In type 1 AGNs, the H$\alpha$ (H$\beta$) emission is dominated by the BLR, whose velocity width is typically much larger than that expected for outflows. 
Any outflow-related contribution would therefore be strongly blended with the BLR emission and cannot be reliably separated. 
Instead, forbidden lines such as \oiii, which arise purely from the NLR, provide a more robust tracer of ionized outflows. 
Consequently, the outflow properties in this work are derived from the \oiii\ profile, while H$\alpha$ (H$\beta$) is modeled using only narrow and BLR broad components.
Then, the emission-line parameters around H$\beta$ (H$\alpha$) can be determined using the Levenberg-Marquardt least-squares minimization technique.
In the middle and right panels of Figure \ref{fig0}, we show two examples of the best fitting results of emission lines around H$\beta$ and H$\alpha$ in our final sample.

Following emission-line fitting, we require:
(1) both H$\alpha$ and H$\beta$ profiles exhibit at least one broad component ($\sigma$ > 600 \kms) with line width and flux three times larger than their uncertainties;
(2) the main narrow lines ([N{\sc ii}], \oiii, and narrow H$\alpha$ and H$\beta$), and the additional broader Gaussian component used to model the [O {\sc iii}] broad component of [O {\sc iii}] have line widths and fluxes greater than three times their uncertainties; (3) both fits around H$\alpha$ and H$\beta$ satisfy $\chi^2/dof<5$.

Some \oiii~emission lines in the parent sample may require three or more Gaussian components for an adequate fit, such as in the case of double-peaked narrow emission lines. However, according to \citet{Zh25}, the fraction of type 1 AGNs with double-peaked narrow emission lines is very small, only about 0.5\%. Moreover, \oiii~profiles with irregular shapes that cannot be well represented by a narrow component plus a single broad component have already been excluded from our sample based on the criterion $\chi^2/dof<5$. Therefore, more complex models for \oiii~profiles are not considered.

\begin{figure*}
\centering\includegraphics[width=8cm,height=5cm]{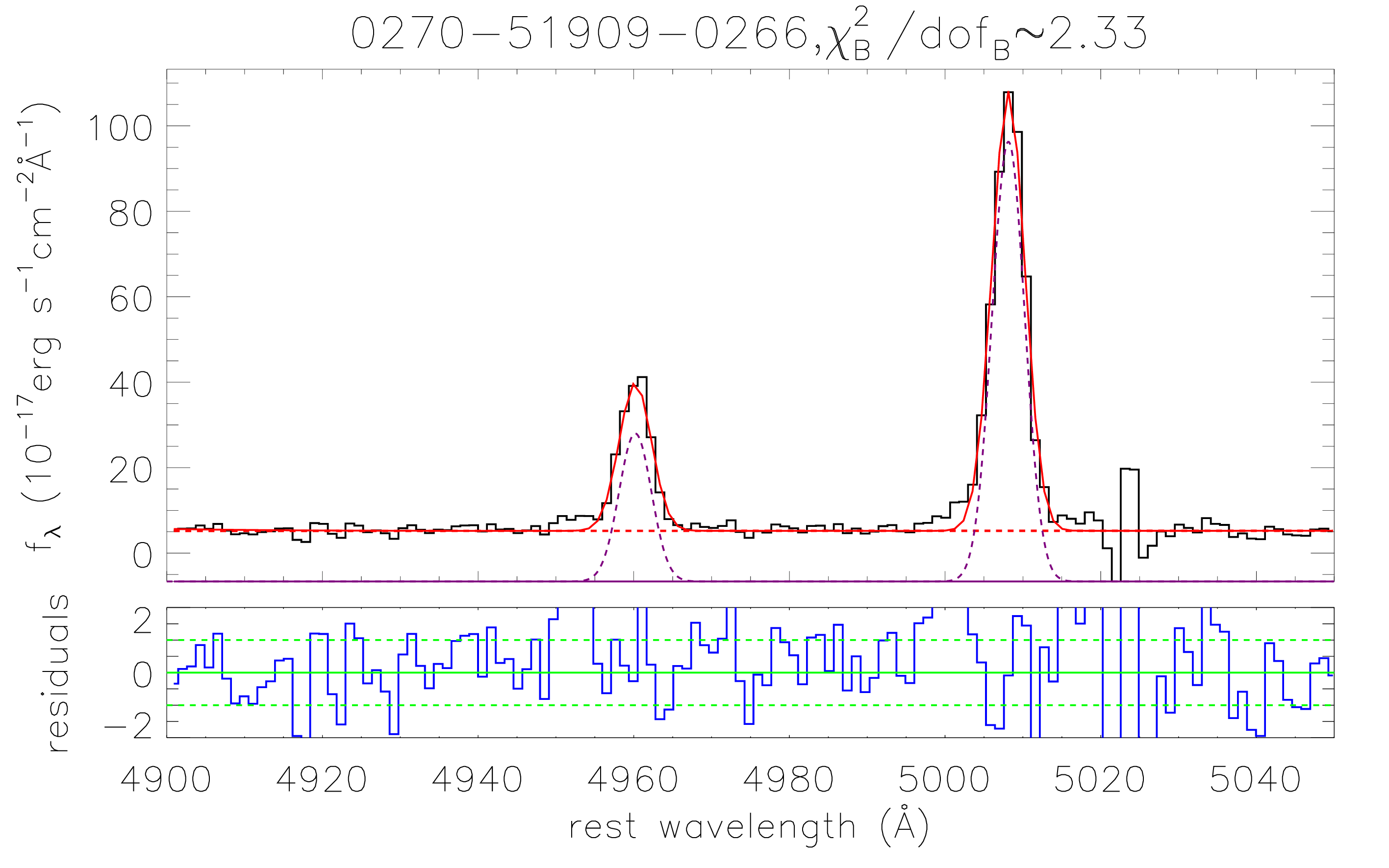}
\centering\includegraphics[width=8cm,height=5cm]{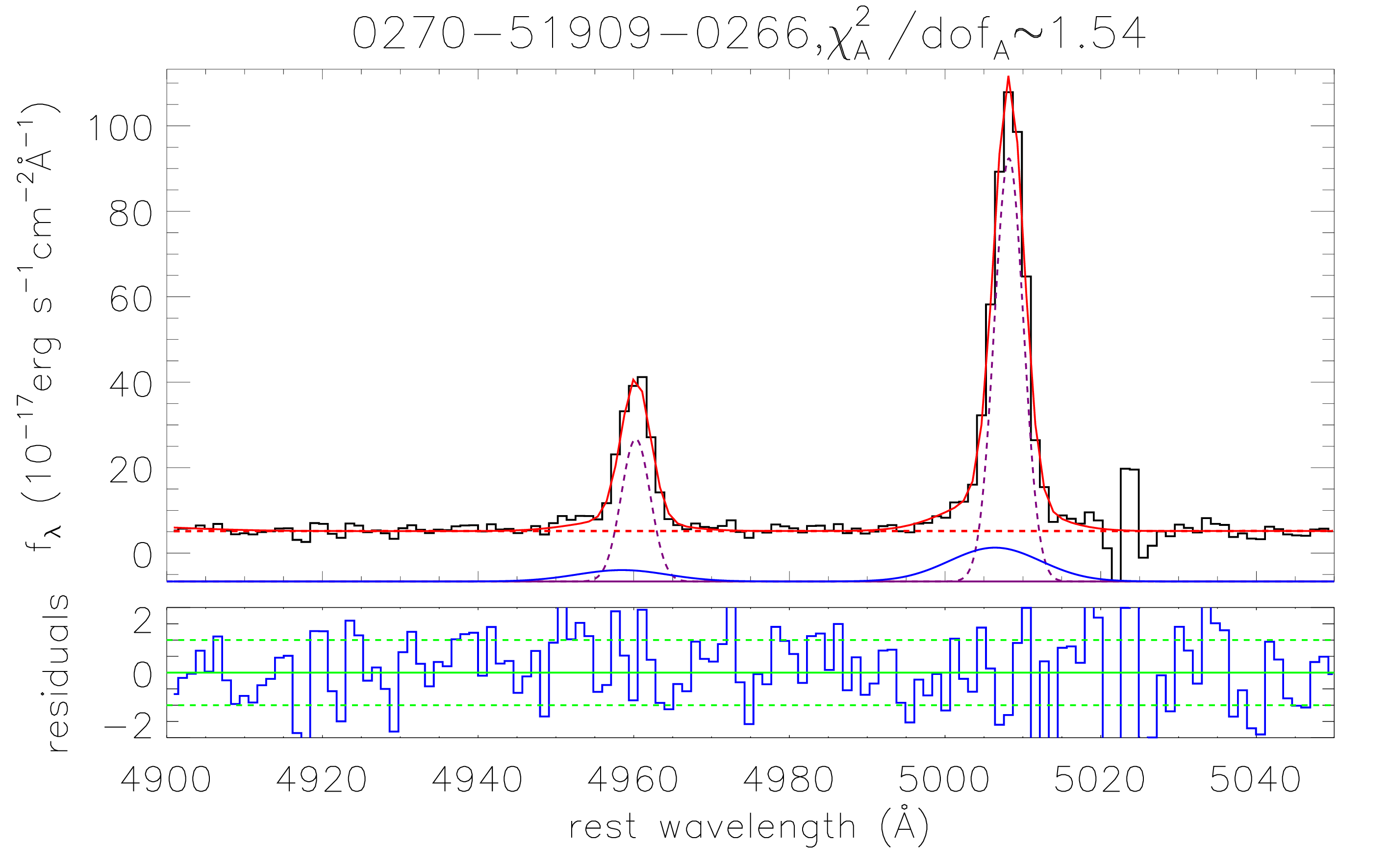}
\caption{An example of type 1 AGN with \oiii~broad component determined by F-test.
In the top panels, the solid black lines show the line spectra after subtracting starlight with Plate-Mjd-Fiberid shown in the title, the solid red lines represent the best fitting results with $\chi_A^2/dof_A$ ($\chi_B^2/dof_B$) for \oiii~doublet with (without) broad component in the title, the dashed purple lines show the core component of \oiii~doublet, and the solid blue lines show the broad component, the dashed red lines represent continuum emission.
In the bottom panels, the solid blue lines represent the normalized residuals, the horizontal solid and dashed green lines show residuals=0, ±1, respectively.
}
\label{f_test}
\end{figure*}

To assess the robustness of the \oiii~broad component, we also perform an F-test to compare two competing models. 
A broad wavelength range is not suitable for performing the subsequent F-test. Therefore, we restrict the analysis to the interval between 4900 and 5050 \AA.
In model A, as described above, the \oiii~doublet is fitted with one narrow and one broader Gaussian component for each of the two \oiii~lines. The corresponding $\chi^2_A$ is re-calculated using the best fitting results with the rest wavelength from 4900 to 5050 \AA, and the degrees of freedom, $dof_A$, are obtained by subtracting eight free parameters from the number of data points (six parameters from the two Gaussian components modeling \oiii~and two from the continuum description).
It is important to note that $dof_A$ does not include the parameters of the optical Fe~{\sc ii} template. This is because the Fe~{\sc ii} emission is treated as a fixed component—its parameters are determined from a broader wavelength fit (4400–5600 \AA) and are held constant in both model A and model B.

In model B, each \oiii~line is modeled using only a narrow Gaussian component, again accompanied by a power-law continuum. All constraints applied to model A are consistently imposed on model B as well. Using the same MPFIT routine, we obtain the best-fitting parameters and compute $\chi^2_B$ and $dof_B$.
With the two sets of $\chi^2$ and $dof$ values, we calculate the F-statistic \citep{Ma97,Ge12,Zh25} as $F_p=(\chi^2_B-\chi^2_A)/(dof_B-dof_A)/(\chi^2_A/ dof_A)$.
Given $dof_B-dof_A$=3, the corresponding statistical F-test shows that an $F_p$ value of approximately 5 corresponds to a significance level of roughly 3$\sigma$. This threshold allows us to evaluate whether the inclusion of an additional broad component in model A is statistically justified over the simpler model B.

Here, we present an example from our final sample (Plate–MJD–FiberID: 0270-51909-0266), shown in Figure \ref{f_test}, to illustrate the reliability of detecting the \oiii~broad component. For this object, the F-test indicates a preference for including a broad component at a confidence level exceeding 5$\sigma$. The two models yield $\chi^2_A = 190$ with $dof_A = 123$ for the model including both core and broad components, and $\chi^2_B = 294$ with $dof_B = 126$ for the model including only the core component.
Applying all the selection criteria above to the full dataset, we identify a final sample of 2,290 type 1 AGNs that exhibit a statistically significant \oiii~broad component.

\section{Main Results}
In this section, basic properties of the 2,290 type 1 AGNs in our sample are measured.

\subsection{main narrow emission lines}

\begin{figure*}
\centering\includegraphics[width = 8cm,height=6cm]{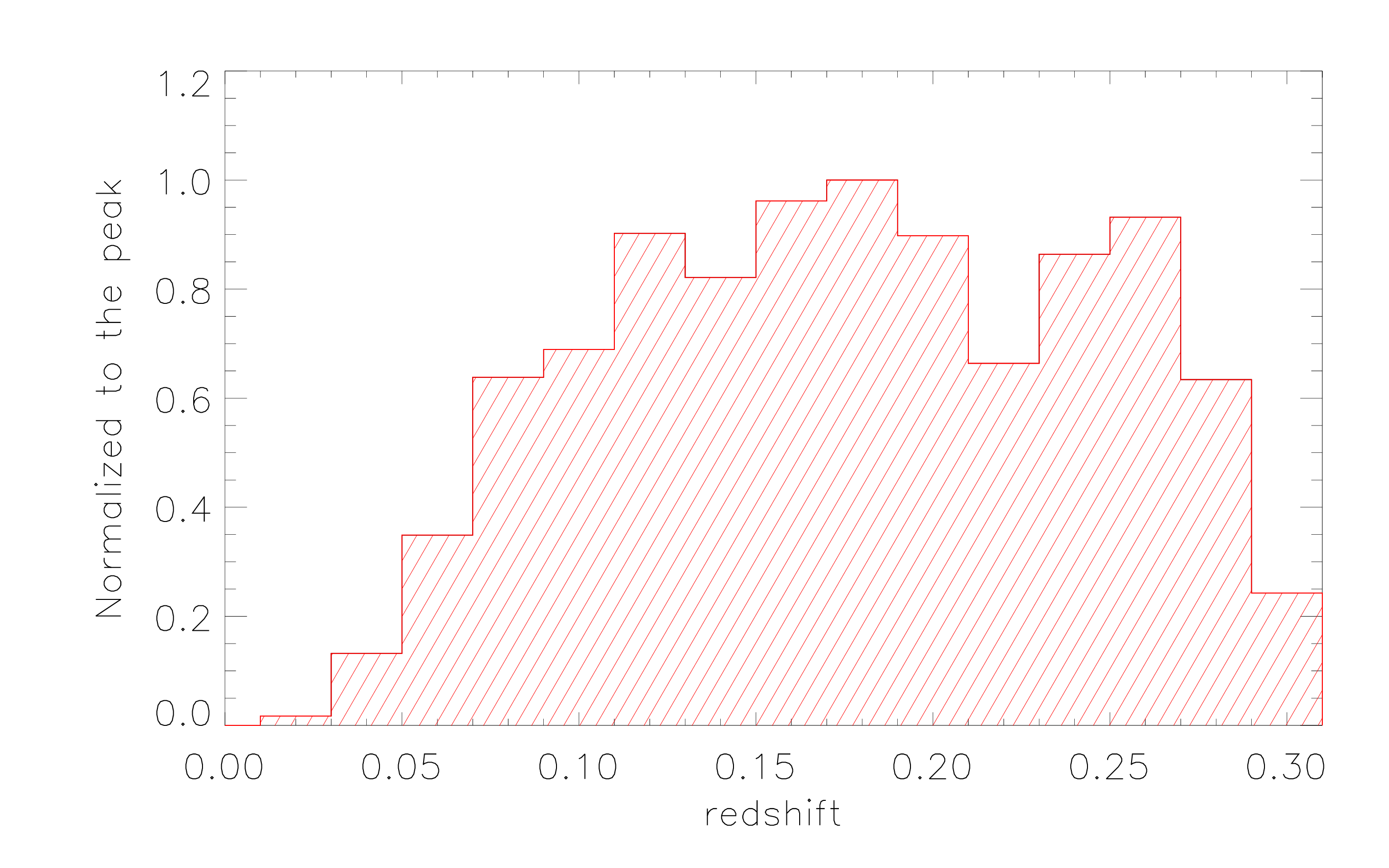}
\centering\includegraphics[width = 8cm,height=6cm]{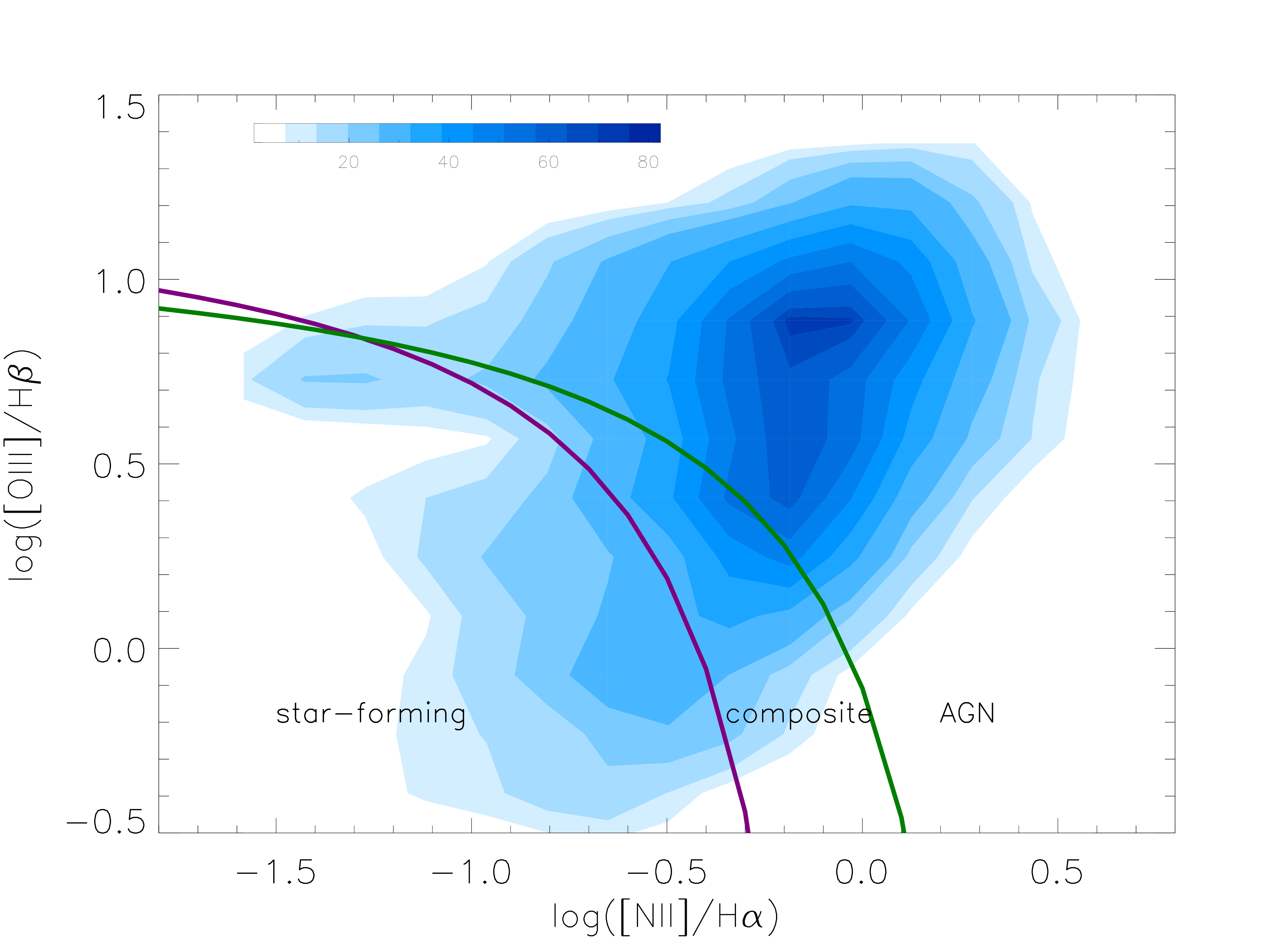}
\caption{The redshift distribution (left panel) and BPT diagram (right panel) for the 2,290 objects in our sample. 
In the right panel, the star-forming, composite, and AGN regions are labeled.
The solid purple line represents the dividing line defined by \citet{Ka03}, below which objects are classified as star-forming galaxies. 
The solid green line shows the dividing line from \citet{Ke01}; objects above this line are classified as AGNs, while those between the two lines are considered composite galaxies.
}
\label{fig1}
\end{figure*}

Figure \ref{fig1} displays the basic characteristics of our sample, with the left panel showing the redshift distribution and the right panel presenting the Baldwin-Phillips-Terlevich (BPT; \citealt{Ke01,Ka03,Ch25}) diagnostic diagram. The sample spans a redshift range of z = 0.013-0.299, with a mean redshift of 0.178.
The BPT diagram in the right panel of Figure \ref{fig1} utilizes the line fluxes obtained from our spectral fitting in Section 2, where the \oiii~fluxes correspond to the narrow-line components. As anticipated, the majority of galaxies in our sample ($\sim$68.21\%; 1,562 objects) fall within the AGN region of the diagram, consistent with our selection criteria of AGNs.
Of the total sample, 17.55\% (402 objects) are located in the star-forming region and 14.24\% (326 objects) in the composite region of the diagnostic diagram.
The ionization of NLR may have a composite origin, with contributions from both AGN activity and star formation, as suggested by \citet{Ro03,Sc21}.

\begin{figure*}
\centering\includegraphics[width =18cm,height=6cm]{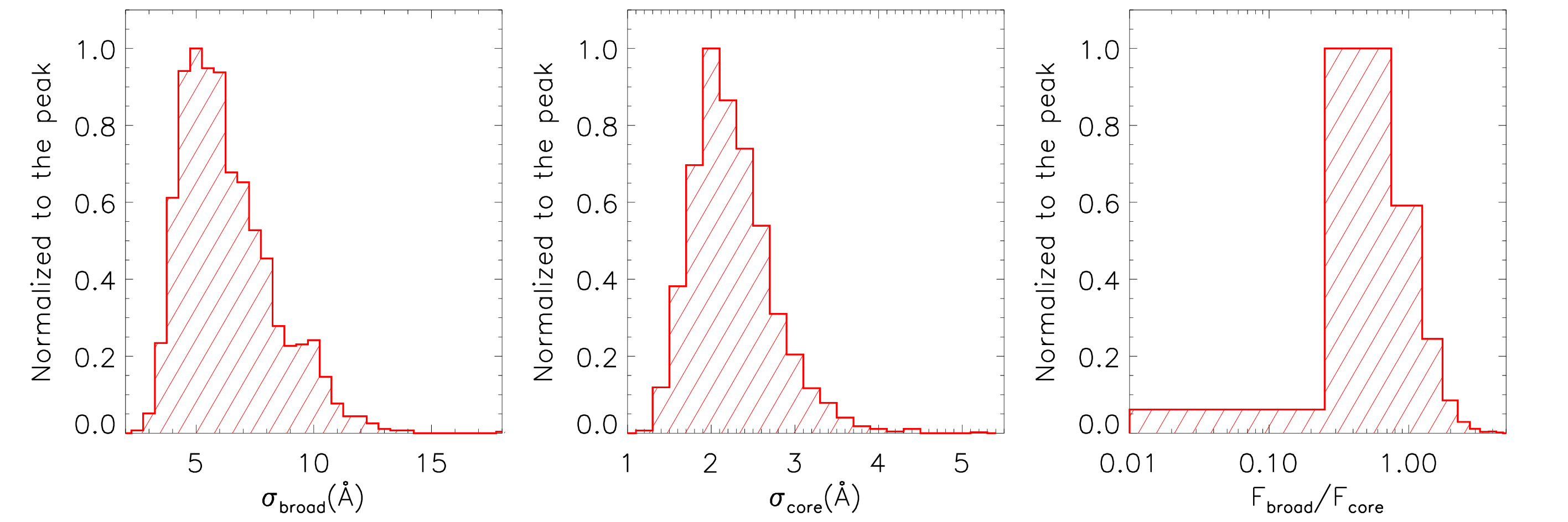}
\caption{The line width $\sigma$ distributions of \oiii~broad (left panel) and \oiii~core (middle panel), and the line flux ratio distribution (right panel).
}
\label{fig2}
\end{figure*}

In the left and middle panels of Figure \ref{fig2}, we present the line width $\sigma$ distributions of the \oiii~core and broad components. 
The $\sigma$ of the \oiii~core ranges from 76.07 to 315.68 \kms. 
The mean $\sigma$ of the \oiii~core is 134.77 \kms, with a standard deviation of 28.75 \kms.
For the \oiii~broad component, the $\sigma$ ranges from 157.54 to 1068.04 \kms, in agreement with earlier studies on the broad component of \oiii~lines in type 1 AGNs \citep{Bi05,Za16}. The broad component has a mean $\sigma$ of 380.37 \kms and a standard deviation of 116.81 \kms.
The distribution of the line flux ratio (\oiii~broad/\oiii~core) is presented in the right panel of Figure \ref{fig2}, spanning a range from 0.20 to 4.56.
Given its asymmetric distribution, we consider the median rather than the mean value here.
The median value of the ratio is 0.72 with a standard deviation of 0.55.
It is observed that although the galaxies exhibit prominent broad features in \oiii, the core component of the emission is generally more pronounced.

\subsection{black hole mass}
To characterize the central engines of the type 1 AGNs in our sample, we estimate their black hole masses using the virialization assumption. In these systems, broad emission lines originate from the rapidly moving gas in the BLR, whose dynamics are predominantly governed by the gravitational potential of the SMBH.
Under the assumption that the BLR gas is virialized, the black hole mass can be inferred from the combination of the BLR size, typically obtained through empirical radius–luminosity relations \citep{Be13}, and the velocity width of the broad emission lines. This approach provides a practical and widely adopted means of estimating black hole masses for large AGN samples.
The virial black hole mass \citep{Pe04,Gr05b,Me22} can be estimated based on broad H$\alpha$ measurement through the equation (2) in \citet{Me22}, and similar method has been applied in our previous work \citep{Zh25}.

\begin{figure*}
\centering\includegraphics[width =18cm,height=9cm]{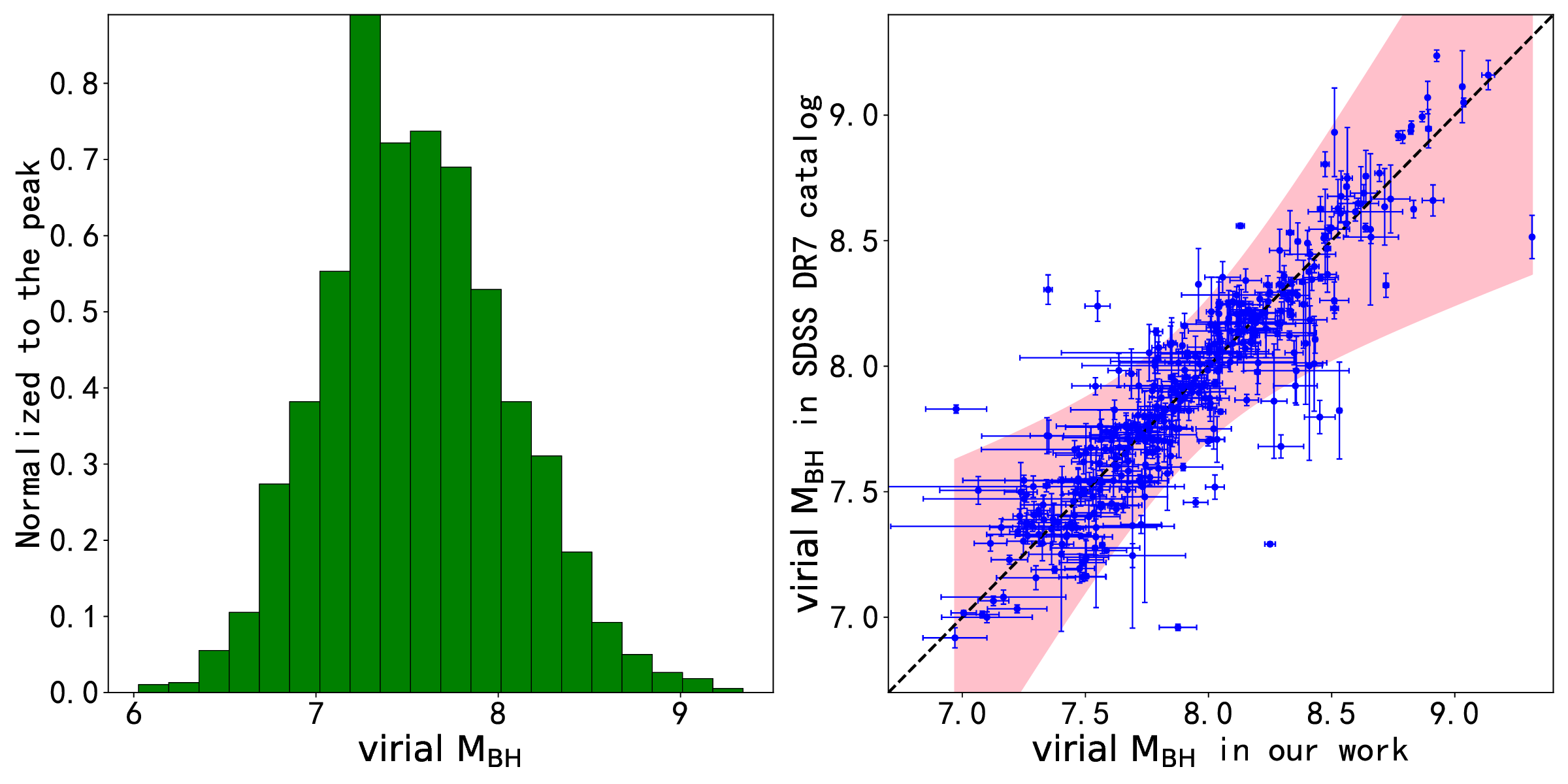}
\caption{The distribution of black hole mass log($\rm M_{\rm BH}$/$\rm M_{\odot}$) for the 2,290 type 1 AGNs in our sample (left panel), and the relation of black hole mass log($\rm M_{\rm BH}$/$\rm M_{\odot}$) for 367 objects estimated in our sample and collected from the SDSS DR7 QSO catalog presented by \citet{Sh11} (right panel). The dashed black line indicates the 1:1 relation. The pink shadow shows 5$\sigma$ confidence bands derived from F-test.
}
\label{bh_dr7}
\end{figure*}

In the left panel of Figure \ref{bh_dr7}, we show the virial black hole mass distribution for the type 1 AGNs in our sample.
The typical uncertainty of these estimates is about 0.13 dex, arising from both the intrinsic scatter of the adopted relation and the measurement uncertainty of line width and luminosity determined through emission-line fitting. For the type 1 AGNs in our sample, the resulting black hole masses span $\log(M_{\rm BH}/M_{\odot}) = 6.03$–$9.34$, with a median value of 7.52 and a standard deviation of 0.50.

To assess the reliability of our virial black hole mass estimates, we cross-match the 2,290 objects in our sample with the SDSS DR7 QSO catalog presented by \citet{Sh11}. A total of 367 objects are successfully matched. 
Although virial black hole masses based on the broad H$\alpha$ line are not provided in the DR7 catalog, the catalog does include the full width at half maximum $FWHM\_BROAD\_HA$ and luminosity of the broad H$\alpha$ component $LOGL\_BROAD\_HA$. For consistency, we derive the virial black hole masses for the DR7 sample using the same equation as employed in this work.
In the right panel of Figure \ref{bh_dr7}, we present a direct comparison between the virial black hole masses derived from our H$\alpha$-based measurements and those listed in the SDSS DR7 catalog.
The mean virial black hole mass obtained in this work is log($\rm M_{\rm BH}$/$\rm M_{\odot}$) = 7.92 with a standard deviation of 0.44, whereas the SDSS DR7 QSO catalog reports a mean value of log($\rm M_{\rm BH}$/$\rm M_{\odot}$) = 7.91 with a standard deviation of 0.45. Overall, the two sets of virial mass estimates exhibit good agreement, demonstrating that our virial black hole mass measurements are broadly consistent with those derived from H$\beta$-based calibrations.

Besides, SMBHs play a pivotal role in active galaxy studies by governing nuclear kinematics and underpinning our understanding of the physical mechanisms operating in the inner regions of AGNs.
It is well established that the stellar velocity dispersion in galactic bulges $\sigma_{\ast}$ correlates with the mass of the central black hole $M_{\rm BH}$. This empirical relationship, commonly referred to as the $M_{\rm BH}-\sigma_{\ast}$ relation \citep{Fe00,Ge00,Ko13}, has been extensively used to estimate black hole masses.
The discovery of the $M_{\rm BH}-\sigma_{\ast}$ relation has profound implications.
In massive galaxies, this relation is often interpreted as evidence for co-evolution between the growth of the central black hole and the evolution of its host galaxy \citep{Kr18}.
In the low-mass regime, it provides important constraints on the formation of black hole seeds at high redshift and the efficiency of black hole growth in small galaxies \citep{Ba20}.
A number of studies have found that AGNs follow the same $M_{\rm BH}-\sigma_{\ast}$ relation as quiescent galaxies \citep{Gr04,Gr08,Ko08,Ji22}.

\begin{figure*}
\centering\includegraphics[width =8cm,height=5cm]{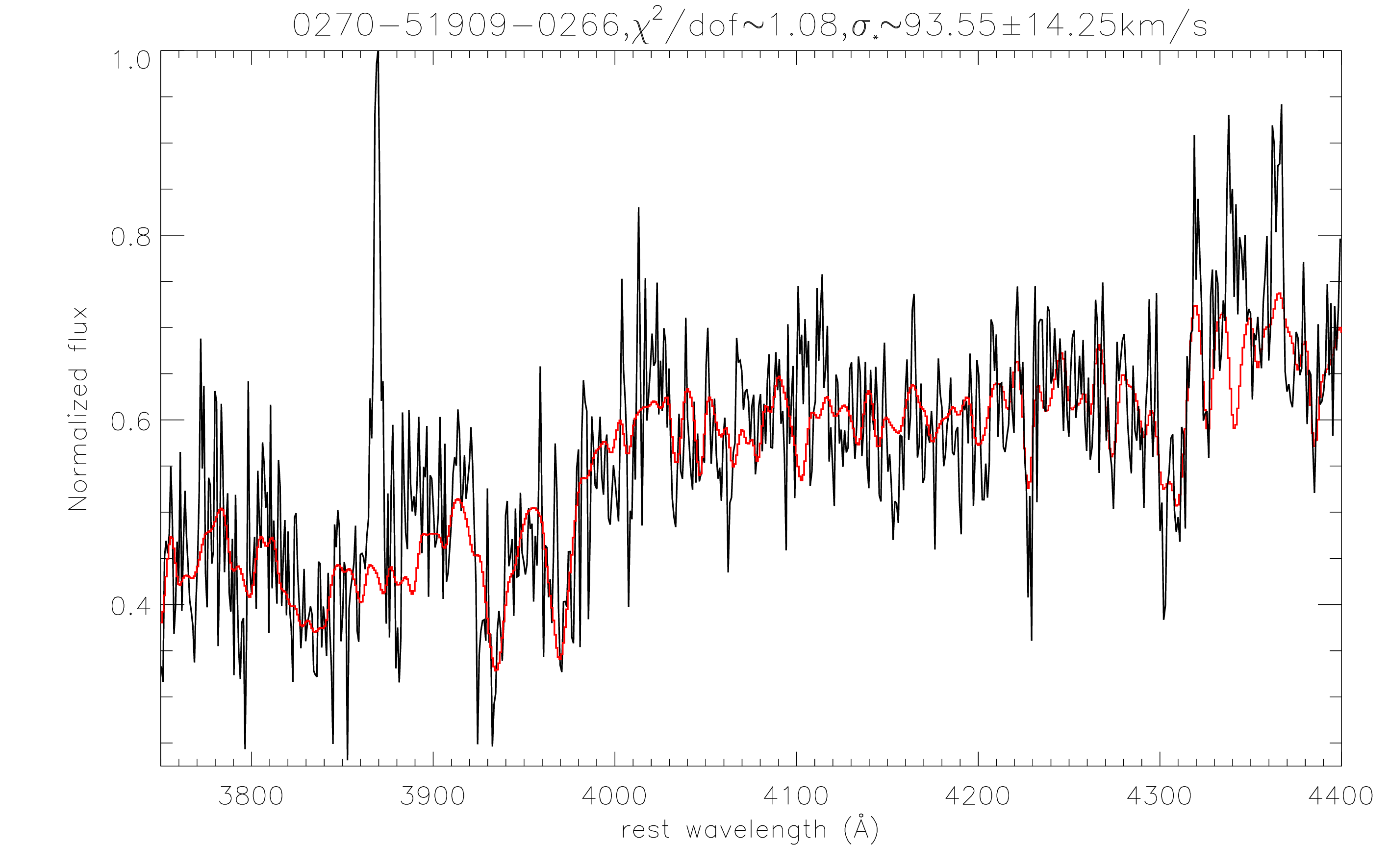}
\centering\includegraphics[width =8cm,height=5cm]{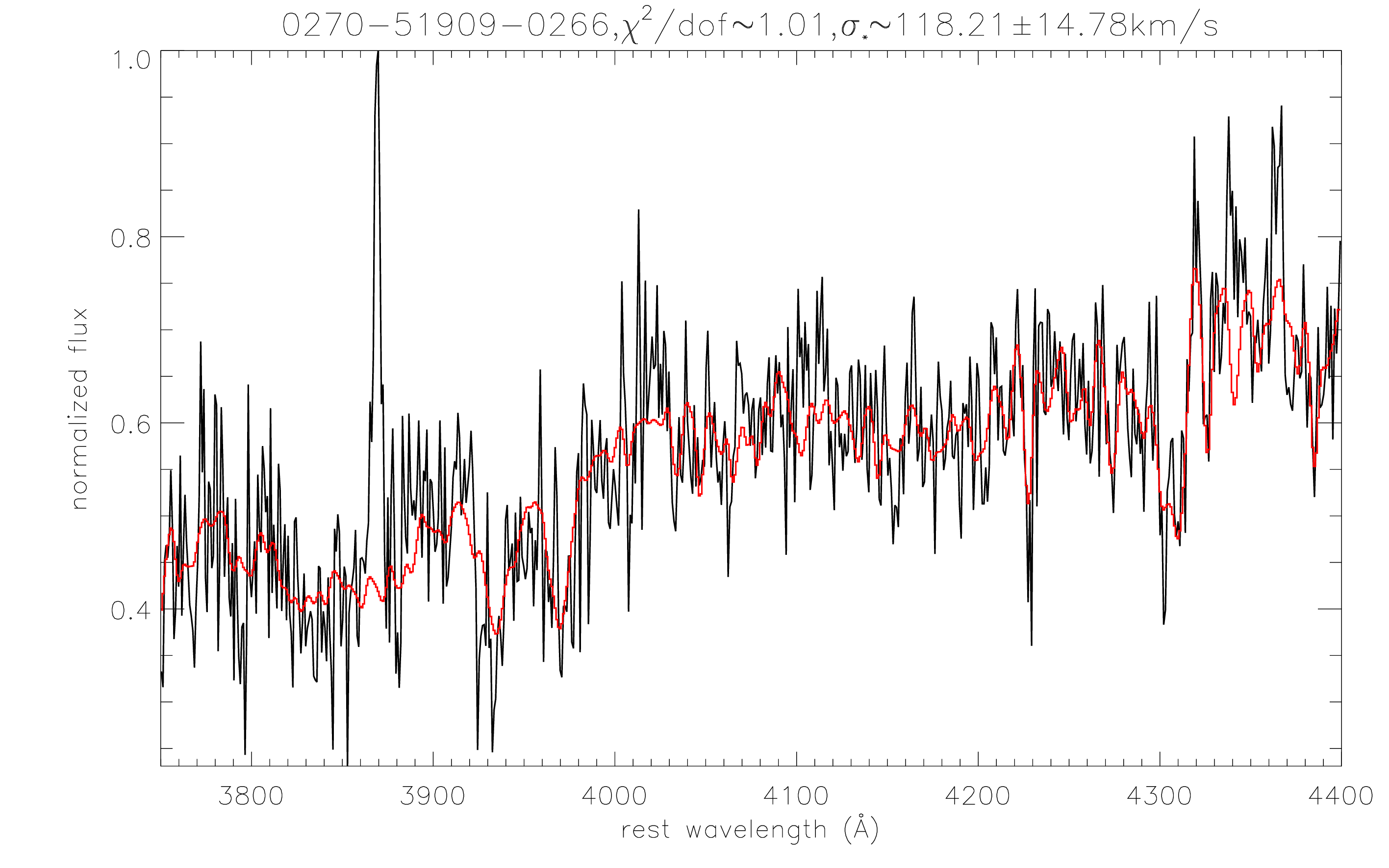}
\caption{An example of object with absorption lines described by the SSP method (left panel) and the direct fitting method (right panel). In each panel, the solid black line shows observed spectrum, the information of Plate-Mjd-Fiberid is shown in the title, the solid red line shows the best fitting results with $\chi^2/dof$ and corresponding $\sigma_{\ast}$ shown in the title.	
}
\label{abs}
\end{figure*}

Although the host-galaxy stellar features in most type 1 AGNs are heavily diluted by the bright nuclear continuum, a non-negligible fraction of objects in our sample still show sufficiently prominent absorption features to allow meaningful stellar kinematic measurements. 
For our final sample of ~2,290 type 1 AGNs, we employ two complementary approaches to measure stellar velocity dispersions $\sigma_{\ast}$ in order to evaluate the robustness of the results obtained from absorption-line fitting.
We apply the SSP method across the full usable wavelength range, which maximizes the information content from multiple absorption features and improves constraints on the stellar continuum.
Besides, we additionally adopt the direct fitting method \citep{Ba02} within a more restricted spectral interval of 3750–4400 \AA. This region contains the prominent Ca H and K absorption features, which are among the most reliable and widely used indicators of stellar kinematics, particularly in type 1 AGNs where the host galaxy contributes only modestly to the total flux. By independently fitting this narrower region, where stellar features are strong and emission line interference is little, we can mitigate systematic biases that may arise in broad-range SSP fitting and provide an essential cross-check on the consistency and reliability of our $\sigma_{\ast}$ measurements.
To ensure robustness, we regard the stellar velocity dispersion as reliable only when the results derived from the full-range fit and the Ca H and K fit agree within 50 \kms.
After applying this consistency criterion, we obtain a subsample of 238 objects with trustworthy $\sigma_{\ast}$ measurements.
In Figure \ref{abs}, we present an example demonstrating that $\sigma_{\ast}$ obtained from the two fitting methods are generally consistent with each other.

\begin{figure*}
\centering\includegraphics[width =18cm,height=9cm]{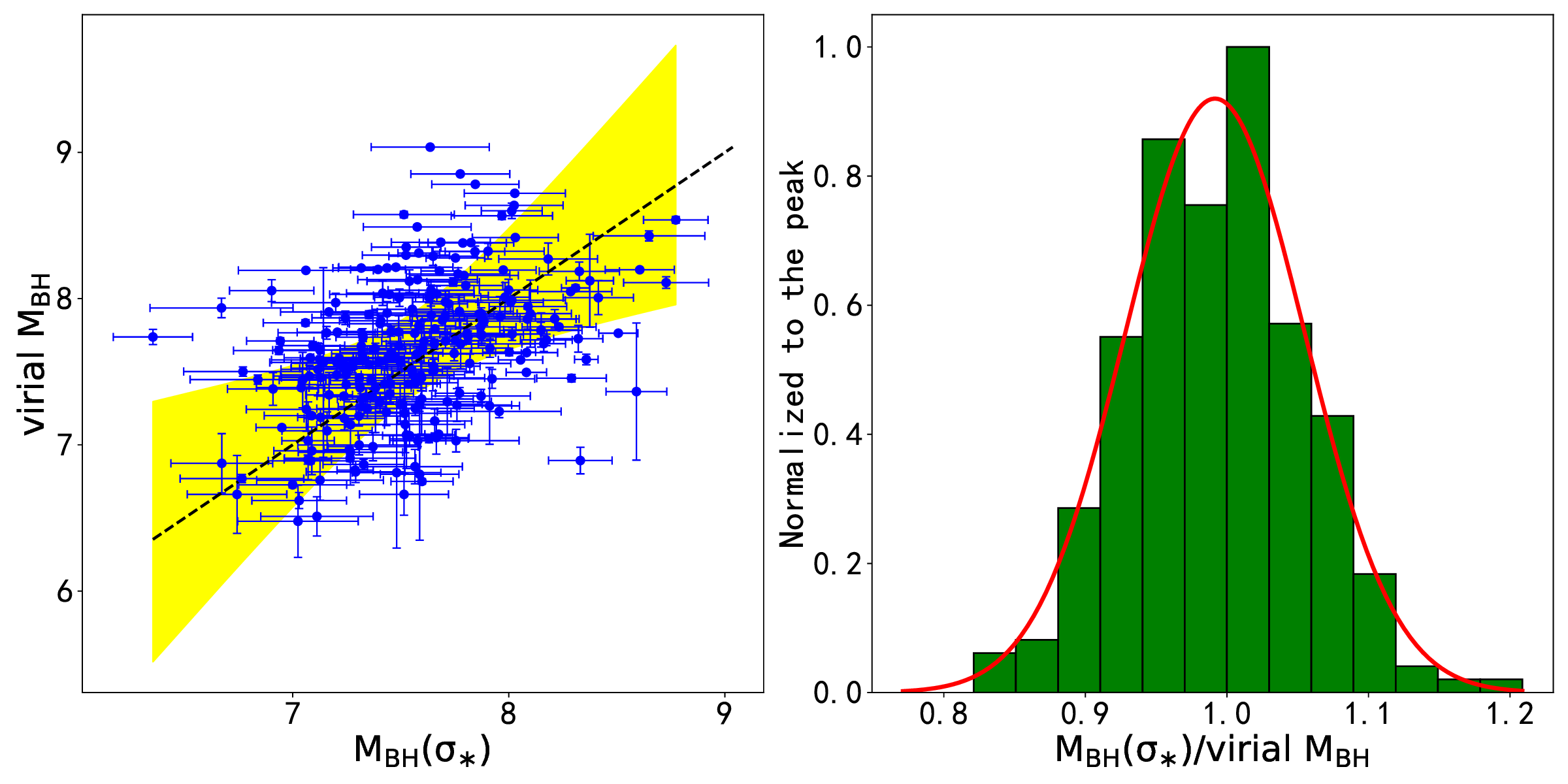}
\caption{The relation of black hole mass log($\rm M_{\rm BH}$/$\rm M_{\odot}$) for 238 objects estimated through the $M_{\rm BH}-\sigma_{\ast}$ relation and through the virialization assumption (left panel), and the black hole mass ratio of the two methods (right panel). In the left panel, the dashed black line indicates the 1:1 relation, and the yellow shadow shows 3$\sigma$ confidence bands derived from F-test. In the right panel, the solid red line show the best result of Gaussian fitting.
}
\label{sigma_bh}
\end{figure*}

Similar as done in \citet{Sc21}, the relation from \citet{Tr02} is applied here, and it can be shown as:
\begin{equation}
\begin{split}	
\log(\frac{M_{\rm BH}}{M_\odot})=(8.13\pm0.06)+	
		(4.02\pm0.32)\times \log(\frac{\sigma_{\ast}}{200 \rm km/s}).
\end{split}
\end{equation}
According to this equation, the black hole mass of the 238 objects can be determined. The left panel of Figure \ref{sigma_bh} illustrates a comparison between the black hole masses of the 238 objects derived from the $M_{\rm BH}$–$\sigma_{\ast}$ relation and those obtained under the virialization assumption. In the right panel of Figure \ref{sigma_bh}, it presents the distribution of the mass ratios between the two approaches. This comparison enables us to directly assess the level of agreement. The mean value of the black hole mass ratio is 0.99 with a standard deviation of 0.06.
Overall, the two methods show a generally good level of consistency, with the majority of objects lying within the 3$\sigma$ confidence bands and the mass ratio distribution centered close to unity. This result suggests that the virialized black hole mass estimates used in this work are broadly reliable and exhibit no significant systematic deviation from those inferred through the $M_{\rm BH}$–$\sigma_{\ast}$ relation.

\subsection{emission profile of~ \rm\oiii}
Based on the \oiii~emission-line fitting described in Section 2, a double-Gaussian decomposition into core and broad components is required for 2,290 type 1 AGNs in our sample. For these objects, we quantify the broad component shift relative to the core $\Delta\upsilon$ by measuring the centroid difference between the core and broad components, defined as $(\lambda_{\rm broad}-\lambda_{\rm core})/5008.24\times3\times10^5$ \kms, where $\lambda_{\rm core}$ and $\lambda_{\rm broad}$ denote the central wavelengths of the core and broad components, respectively. Using this criterion, we find that 1,922 objects exhibit significant blue-shifted broad components ($\Delta\upsilon<-50$ km s$^{-1}$), 290 show nearly symmetric broad components ($-50<\Delta\upsilon<50$ km s$^{-1}$), and 78 display red-shifted broad components ($\Delta\upsilon>50$ km s$^{-1}$).

\begin{figure}
\centering\includegraphics[width =9cm,height=5.2cm]{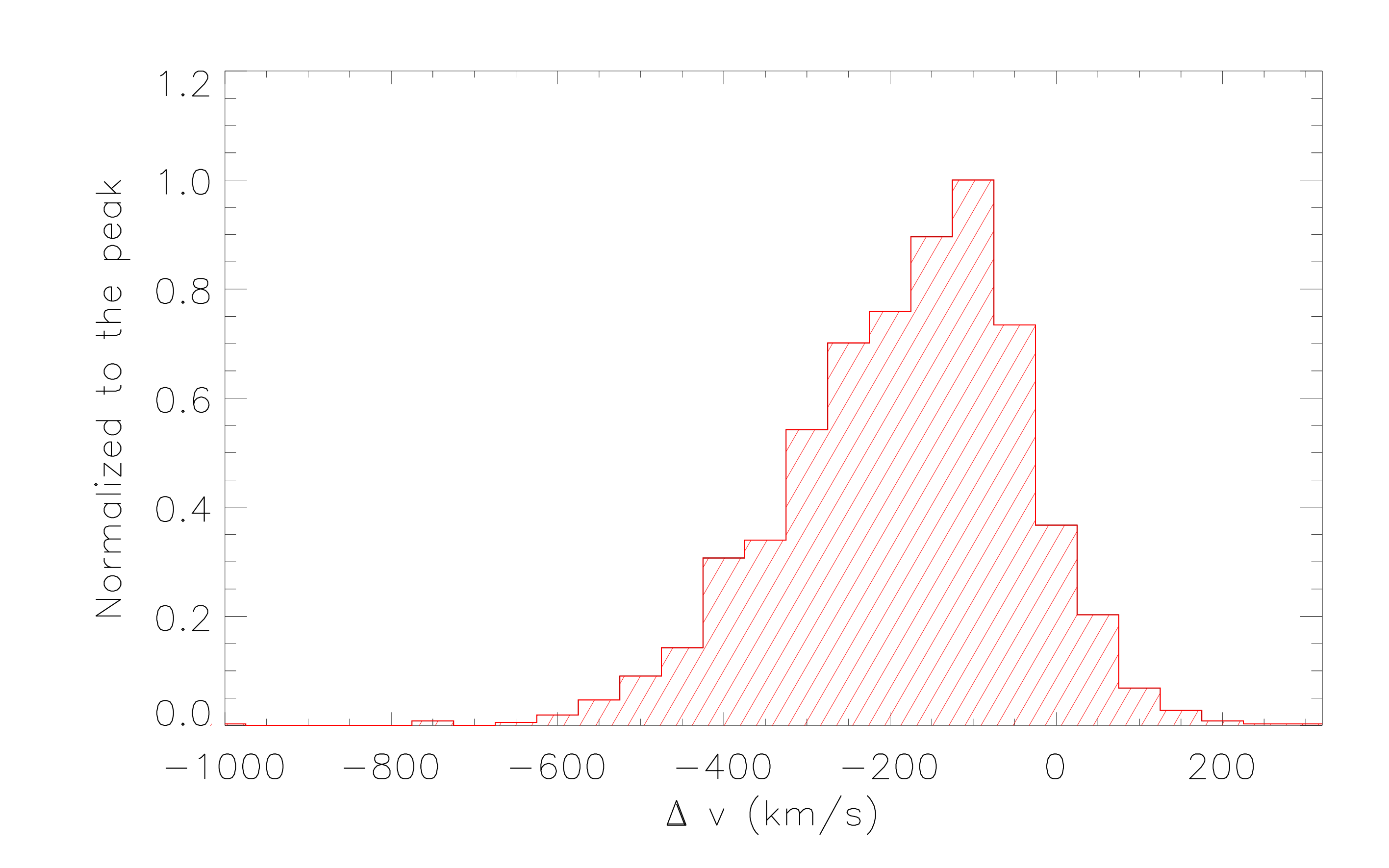}
\caption{The distribution of $\Delta\upsilon$ for the 2,290 type 1 AGNs in our sample.}
\label{fig3}
\end{figure}

In Figure \ref{fig3}, we show the $\Delta\upsilon$ distribution of the 2,290 type 1 AGNs with \oiii~broad component in our sample.
The distribution of $\Delta\upsilon$ in the sample is characterized by a mean value of -180.86 \kms~with a standard deviation of 139.87 \kms. Moreover, the $\Delta\upsilon$ values range from -976.13 to 309.53 \kms.
For the type 1 AGNs in our sample that exhibit broad components, we further examine the relationship between $\Delta\upsilon$ and the line width $\sigma$ of the \oiii~core. There is a weak correlation between these two parameters with a Pearson coefficient as -0.33 and a p-value of less than $10^{-10}$. This indicates that objects with broader core line widths tend to display more blue-shifted broad components, consistent with previous findings \citep{Bi05,Sc21}.

\begin{figure}
\centering\includegraphics[width =8cm,height=6cm]{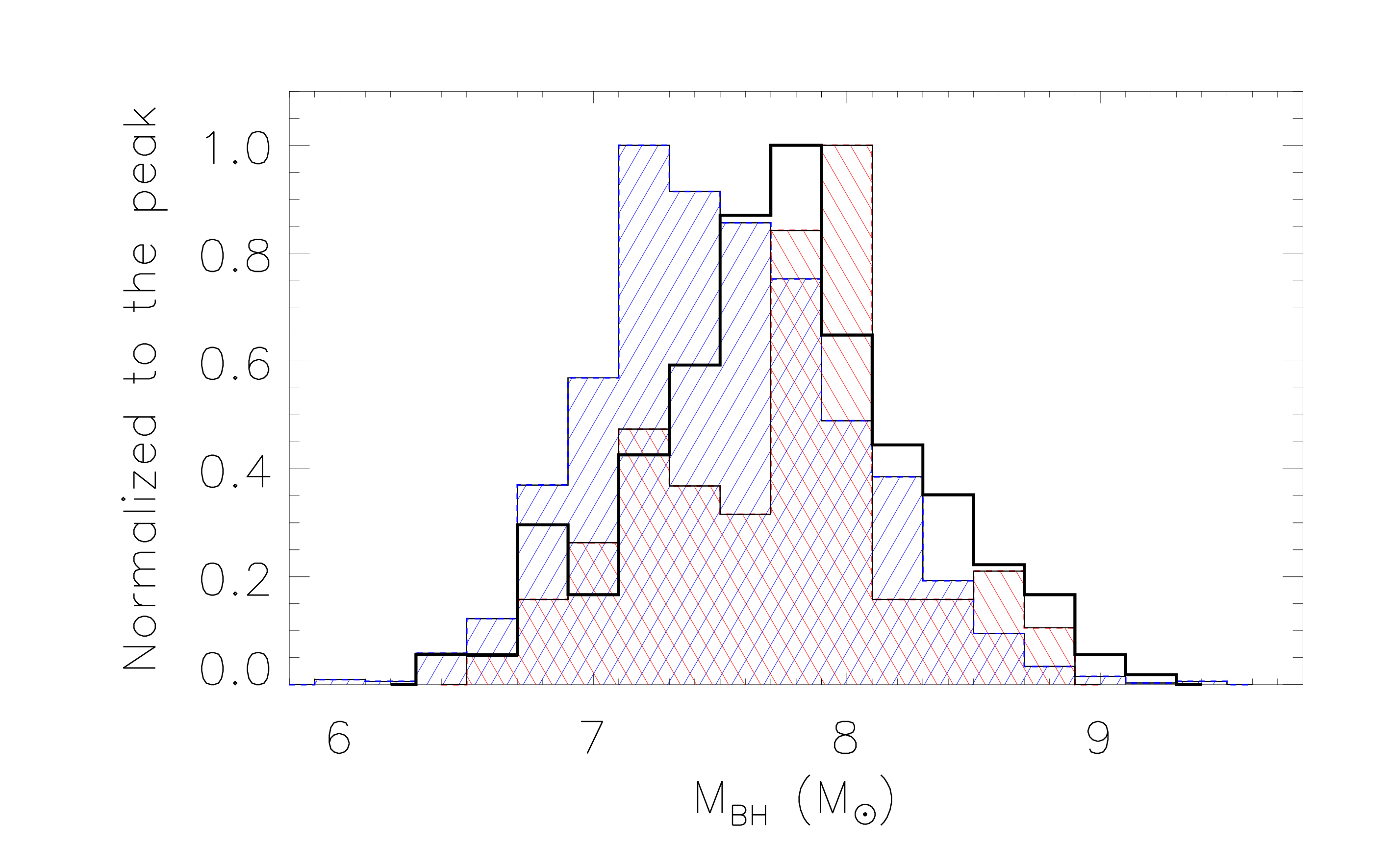}
\caption{Black hole mass distribution for objects with blue-shifted broad components (histogram filled with blue line), with red-shifted broad components (histogram filled with red line), and with symmetric \oiii~profiles (solid black line).
}
\label{fig5}
\end{figure}

 
Up to this point, [O~{\sc iii}] lines exhibiting a broad component have been identified in AGNs spanning a wide range of black hole masses.
To explore whether the observed kinematic appearance of the outflow depends on black hole mass, we compare the black hole mass distributions among AGNs with different [O~{\sc iii}] profile morphologies: (1) blue-shifted asymmetric profiles, (2) symmetric profiles, and (3) red-shifted asymmetric profiles.
All three subsamples are consistent with the presence of outflows. The observed asymmetries likely arise partly from orientation and obscuration effects, where the receding side of the outflow may be hidden by the host galaxy disk or circumnuclear dust. Therefore, this comparison does not test the presence of outflows, but rather whether black hole mass influences their observable kinematic signatures.
Figure \ref{fig5} presents the black hole mass distributions for these three populations, which appear to be broadly similar.
AGNs exhibiting blue-shifted broad components have a mean black hole mass of log($\rm M_{\rm BH}$/$\rm M_{\odot}$) = 7.51 with a standard deviation of 0.49, spanning a range from 6.03 to 9.34. For AGNs with symmetric \oiii~lines, the distribution shows a mean of 7.73 and a standard deviation of 0.49, with values ranging from 6.56 to 8.89. Meanwhile, AGNs with red-shifted broad components display a mean of 7.75 and a standard deviation of 0.53, covering the range 6.37–9.16.
The similarity among these distributions indicates that black hole mass does not play a dominant role in determining the observed [O~{\sc iii}] profile shape. Instead, the diversity of line asymmetries is more likely driven by geometric and line-of-sight effects rather than intrinsic differences in black hole mass.

One of the aims of this work is to explore the potential connection between black hole mass and the properties of the \oiii~broad components. To quantify this, we adopt the blue emission parameter, which measures the full extent of the broad component on the blue side for all objects with [O~{\sc iii}] lines fitted by a double-Gaussian decomposition, including those exhibiting blue-shifted, symmetric, or red-shifted broad components. The blue emission is defined as ($\Delta\upsilon - \rm FWHM_{\rm broad}$), where $\rm FWHM_{\rm broad}$ represents the full width at half maximum of the broad component in units of \kms. 
In our sample, the blue emission ranges from -2730.25 to -409.11 \kms, with a mean value of -1015.20 \kms and a standard deviation of 347.56 \kms.

It is important to note that the $\rm FWHM_{\rm broad}$ used in this work is derived from a Gaussian fit to the broad component, whereas \citet{Sc18, Sc21} defined the $\rm FWHM_{\rm broad}$ as half of the total width of the broad component at its base. These two definitions characterize different aspects of the line profile, and therefore may naturally lead to systematic differences in the measured widths and in any physical quantities inferred from them. In particular, the `half-width at the base' statistic is highly sensitive to the precise determination of the line boundaries, which can be significantly affected by noise and continuum placement. As a result, this definition may introduce larger measurement scatter or bias, especially for spectra with moderate signal-to-noise ratios. In contrast, a direct Gaussian fit to the broad component provides a more stable and reproducible estimate of its characteristic velocity scale, since the fitting procedure effectively reduces the sensitivity to ambiguities in defining the line base. For these reasons, we adopt a Gaussian-fit–based $\rm FWHM_{\rm broad}$, which offers better robustness and consistency across our sample.

\begin{figure*}
\centering\includegraphics[width =8cm,height=6cm]{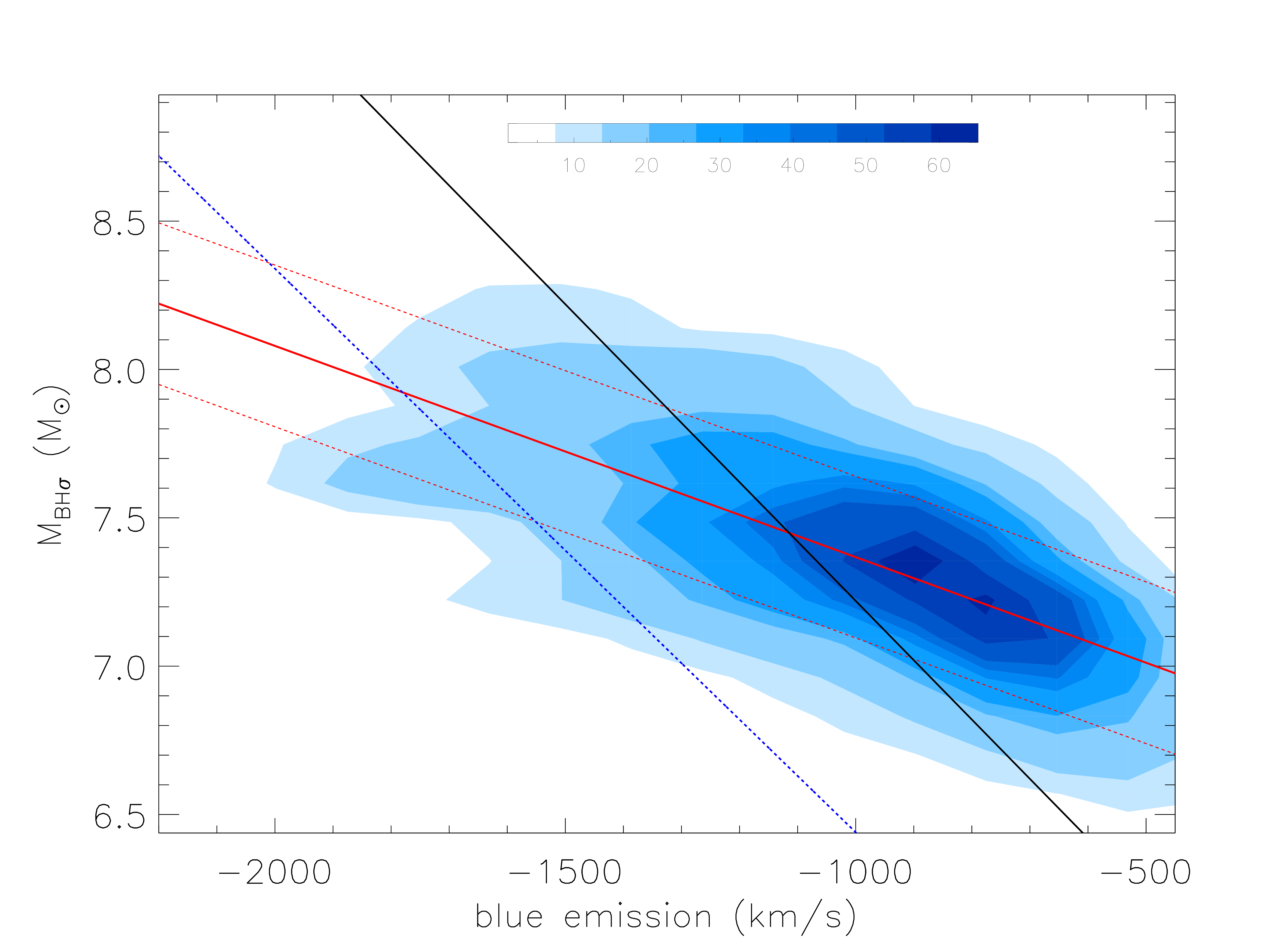}
\centering\includegraphics[width =8cm,height=6cm]{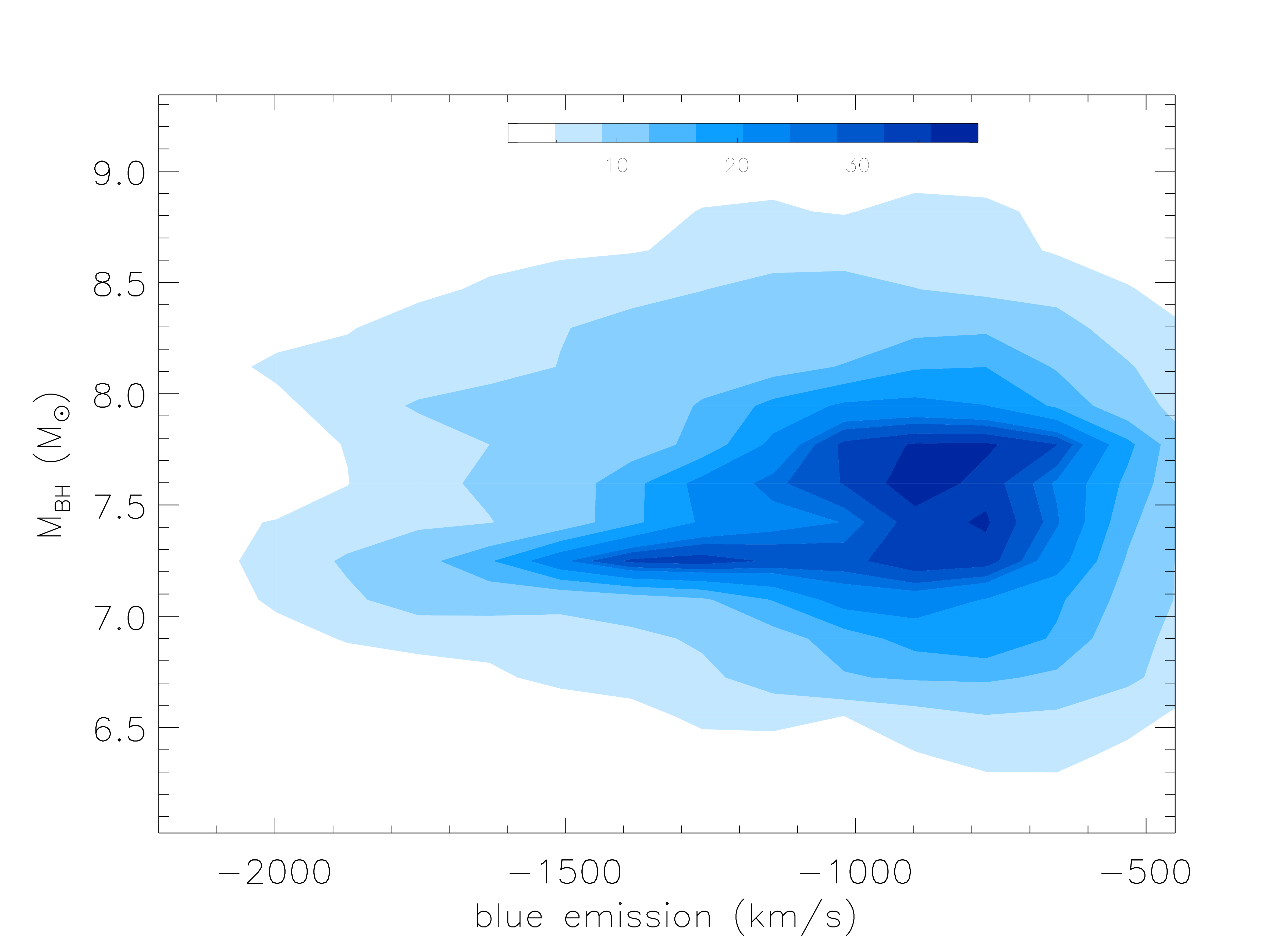}
\caption{The relation between blue emission and black hole mass for the 2,290 type 1 AGNs in our sample.
The left panel shows the black hole mass determined by $M_{\rm BH}-\sigma_{\ast}$ relation, and the right panel shows the virial black hole mass.
In the left panel, the solid and dashed red line represent the best fitting with Linear Least-squares approximation and 1$\sigma$ scatter for our sample, respectively.	
The solid black line and dashed blue lines represent the best fitting for the 22 broad line Seyfert 1 galaxies in \citet{Sc21} and 28 narrow line Seyfert 1 galaxies in \citet{Sc18}, respectively.	
}
\label{fig4}
\end{figure*}

We further investigate the relation between the blue emission parameter and black hole mass. 
Following the methods in \citet{Sc18,Sc21}, the line width $\sigma$ of the \oiii~core component is employed as a proxy for $\sigma_{\ast}$, from which black hole masses $M_{\rm BH,\sigma}$ are derived using the $M_{\rm BH}-\sigma_{\ast}$ relation above. 
For the 2,290 type 1 AGNs in our sample, these two quantities exhibit a strong correlation, with a Pearson coefficient of $-0.65$ and a p-value $<10^{-10}$. The correlation is shown in the left panel of Figure~\ref{fig4}, together with a linear least-squares fit to our data, which yields a slope of $-0.0007$ and an intercept of 6.66 (solid red line). For comparison, the best-fit relation reported by \citet{Sc21} for 22 broad-line Seyfert 1 galaxies has a slope of $-0.0020$ and an intercept of 5.22 (solid black line), while that obtained by \citet{Sc18} for 28 narrow-line Seyfert 1 galaxies has a slope of $-0.0019$ and an intercept of 4.54 (dashed blue line). 
These results seem to indicate that AGNs with \oiii~broader component tend to host more massive black holes, although the slope derived from our sample is noticeably smaller than those reported in \citet{Sc18,Sc21}.

\section{Discussions}
The asymmetric or broadened profiles of \oiii~often trace kinematic disturbances in the NLRs. Numerous studies have demonstrated that powerful outflows is frequently linked to radiation-pressure or magnetic driving mechanisms processes \citep{Po00,Ki12,Ha14}, while the outflows have often been connected with the presence of an asymmetry in the \oiii~lines \citep{Gr05,Be16}. 
In particular, the \oiii~line has proven to be a sensitive tracer of ionized outflows on kiloparsec scales, providing insights into how supermassive black holes interact with their host galaxies \citep{Wy16}. 
Therefore, exploring the properties of \oiii~emission, and especially its asymmetric components, is crucial to disentangling the impact of AGN-driven outflows on galaxy evolution.

In the present work, we investigate a large sample of 2,290 type 1 AGNs from SDSS DR16 that exhibit a broad component in their \oiii$\lambda$5007 emission profile. For our sample, the core component shows line width ($\sigma$) ranging from 76.07 to 315.68 \kms, with a mean value of 134.77 \kms. By contrast, the broad component spans a significantly broader range of 157.54 to 1068.04 \kms, with a mean $\sigma$ of 380.37 \kms. This systematic difference in line width between the core and broad components strongly suggests that the broad components originate from physically distinct regions within the narrow-line region, consistent with earlier findings \citep{Bi05,Sc18,Sc21}.

The broad components exhibit line widths that fall between those of the \oiii~core component produced in the NLRs and the broad Balmer lines originating from the BLRs. 
In many systems, the gas velocity dispersion tends to decrease with increasing distance from the central engine \citep{Sc16}, which suggests that these broad components may be associated with relatively inner regions of the NLR \citep{Sc18}.
This intermediate zone, located between the classical [O~{\sc iii}] core–emitting region and the BLR, appears to be more dynamically perturbed than the outer NLR \citep{Ho03,Bi05}.
However, the radial behavior of outflow kinematics is not necessarily monotonic. Recent integral-field spectroscopic and theoretical studies \citep{Ki15,Ma25} show that outflow velocities can rapid acceleration at approximately 1 kpc. Therefore, the [O~{\sc iii}] broad components should be interpreted more generally as tracing kinematically disturbed outflowing gas within the NLR, rather than being uniquely confined to its innermost regions.

Previous studies indicate that outflows reduce the \oiii~equivalent width while broadening its profile \citep{Cr16}, with the equivalent width further correlated with the blue-shifted broad component \citep{Lu12}. In line with these findings, we also detect a weak connection between $\Delta\upsilon$ and the line width of the \oiii~core component. Furthermore, by quantifying the full extent of the \oiii~broad component relative to the centroid of the core component through the blue emission parameter, we detect a strong correlation with black hole mass $M_{\rm BH,\sigma}$, characterized by a Pearson coefficient of -0.65.
Comparable trends have also been reported by \citet{Sc18,Sc21}, and the correlations are similar in their samples of 28 narrow-line Seyfert 1 galaxies (Pearson coefficient of -0.63; \citealt{Sc18}) and 22 broad-line Seyfert 1 galaxies (Pearson coefficient of -0.79; \citealt{Sc21}). 
It should be noticed that the stellar velocity dispersion here traced by the narrow-line width is used to estimate black hole masses through the $M$–$\sigma_{\ast}$ relation.
However, when adopting virial black hole masses, the correlation between $M_{\rm BH}$ and the blue emission parameter becomes negligible, with a Pearson correlation coefficient of only –0.06, and the results are shown in the right panel of Figure~\ref{fig4}. This discrepancy may arise because, despite the fact that narrow-line widths can in principle serve as proxies for stellar velocity dispersion, the intrinsic scatter between the two quantities is quite large. As a result, black hole masses inferred from the $M$–$\sigma_{\ast}$ relation may introduce substantial statistical biases, thereby producing an artificial correlation with the blue emission parameter.

\begin{figure}
\centering\includegraphics[width =8cm,height=5cm]{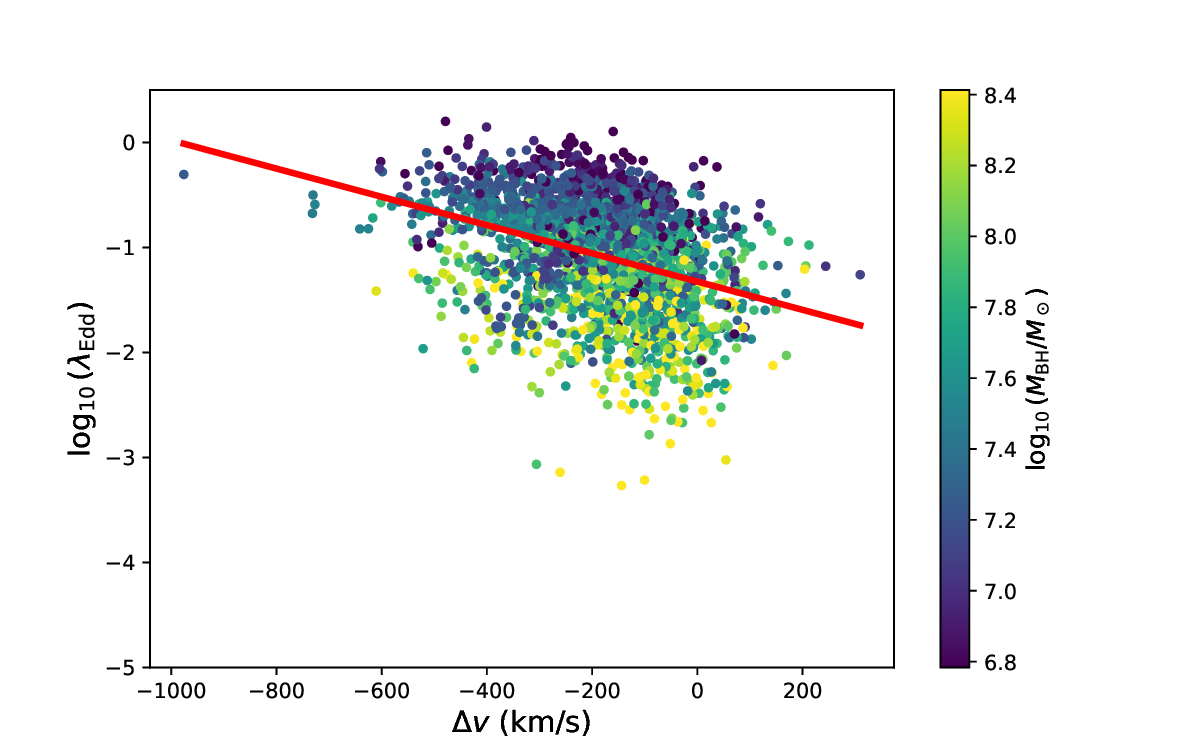}
\caption{The relation between velocity shift and the Eddington ratio for the 2,290 type 1 AGNs in our sample.
The solid red line represents the best fitting result by Linear Least-squares approximation.
}
\label{edd}
\end{figure}

When focusing on the broad component of the \oiii~emission line, it is important to examine the role of the Eddington ratio because the broad component is widely interpreted as tracing outflows. Since radiative pressure from the accretion disk increases with accretion rate, objects with higher Eddington ratios are expected to drive stronger or faster outflows, making the broad component sensitive to the level of accretion activity. Therefore, incorporating the Eddington ratio into the analysis of this sample allows us to assess whether the observed velocity shifts ($\Delta\upsilon$) of both the broad and core components of the \oiii~emission line are regulated by radiation-driven mechanisms.
The Eddington ratio is a main parameter to trace central AGN activities and it can be defined as $\rm \lambda_{Edd}=L_{bol}/L_{Edd}$, where $\rm L_{bol}$ and $\rm L_{Edd}$ ($\rm L_{Edd}=1.26\times10^{38}M_{BH}/M_\odot$) are the bolometric luminosity and the Eddington (maximum) luminosity of material accreting onto a massive black hole with its particular mass, respectively \citep{Al22}. In addition, $\rm L_{bol}$ can be estimated through 5100\AA~based bolometric luminosity \citep{kas00,Sh11,db20,nh20}.
However, it is noteworthy that bolometric correction factor has a considerable range from 8.0 to 23.1 \citep{Ki17}.
Here, the accepted optical bolometric correction factor is 9.8, as commonly applied in the literature \citep{Mc04,Ma04,Ya20}.
As a result, we find that the velocity shifts ($\Delta\upsilon$) between the broad and core components of the \oiii~emission line exhibit a weak correlation with the Eddington ratio, and the Pearson coefficient is -0.35 with a p-value of less than $10^{-10}$.
And this relation is shown in Figure \ref{edd}.

Therefore, objects with higher Eddington ratios tend to show larger shifted broad components.
But the orientation and projection effects can dilute intrinsic velocity signatures. If an outflow has a preferred axis, such as jet-driven outflows, differences in line-of-sight angle will cause otherwise similar flows to appear with widely varying radial velocities.
Besides, the presence of radio jets can induce kinematic disturbances the ionised gas \citep{Mo19,Ku25}. As a result, objects hosting radio jets may exhibit more disturbed kinematics than radio-quiet AGN.
To explore this possibility, we then examine the SDSS–FIRST matched catalog, which provides radio information for SDSS sources with FIRST counterparts. 
Among the 2,290 type 1 AGNs in our sample, 1,739 objects show radio fluxes below the detection threshold (recorded as zero), suggesting that they are radio-quiet AGNs.
For the 1,739 radio-quiet AGNs, the relation between $\Delta\upsilon$ and the Eddington ratio shows a significant correlation, with a Pearson coefficient of -0.32 and a p-value $<10^{-10}$. Therefore, even after accounting for the potential influence of radio jets, the correlation remains essentially the same as that found for the full sample of type 1 AGNs.


To assess whether the presence of ionized outflows is linked to global accretion properties, a control sample is constructed by applying the same selection criteria used for the 2,290 type 1 AGNs with a broad component (outflow sample), except that objects exhibiting a broad component are excluded. In this way, type 1 AGNs without a broad component (non-outflow sample) are retained as the comparison sample. Ultimately, 3,611 type 1 AGNs without a broad component are selected as the control sample.

\begin{figure*}
\centering\includegraphics[width =8cm,height=6cm]{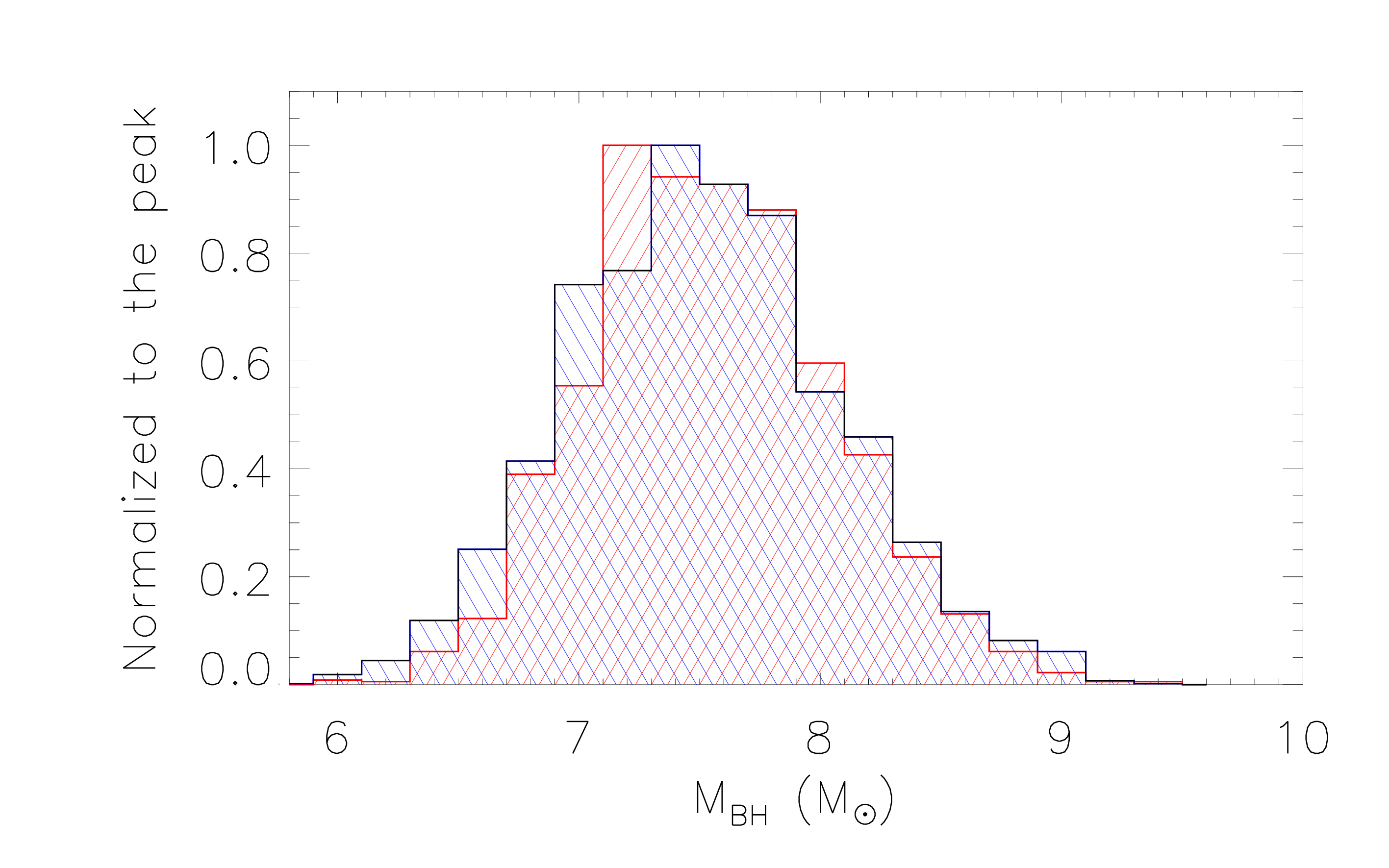}
\centering\includegraphics[width =8cm,height=6cm]{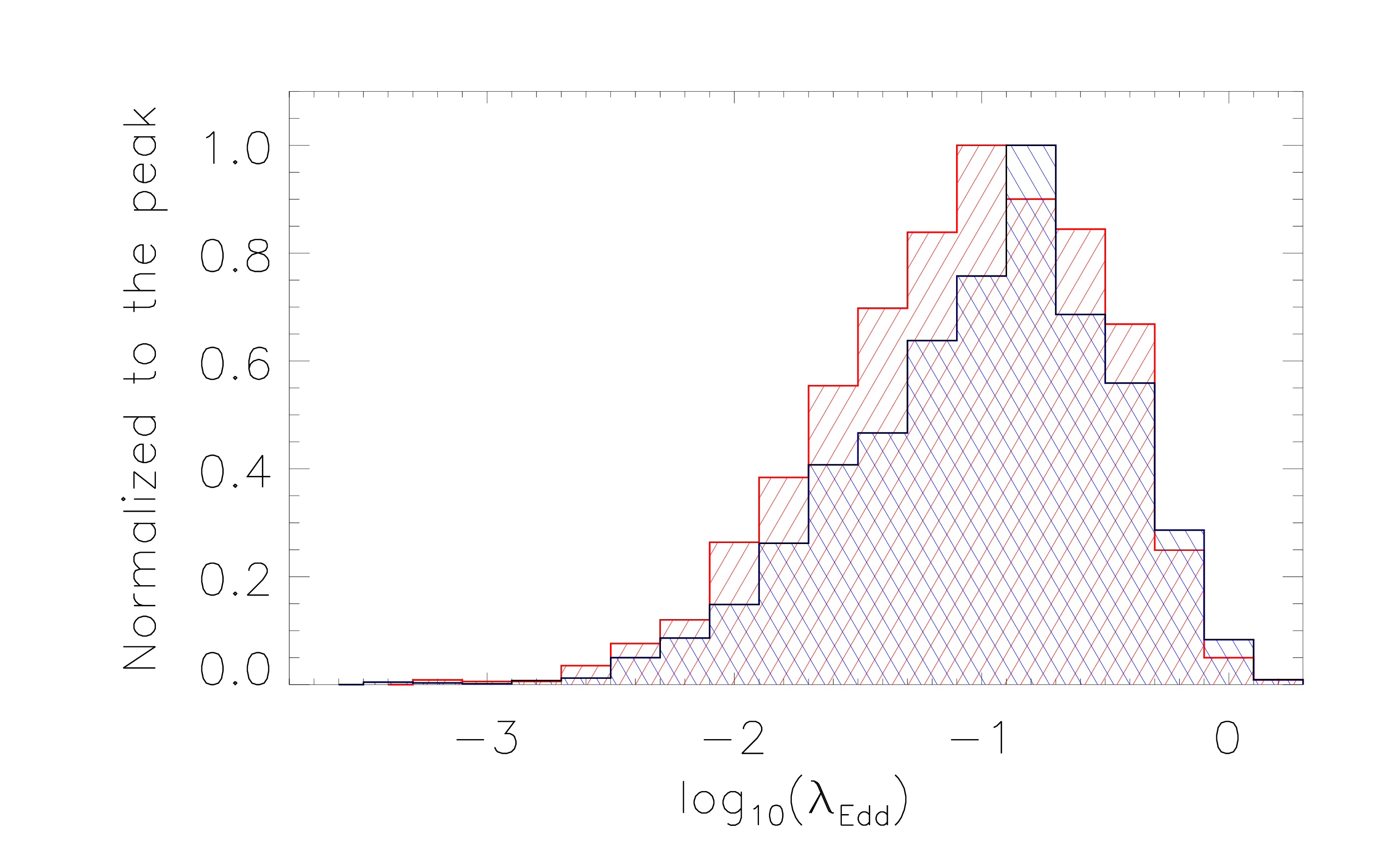}
\caption{The distribution of black hole mass log($\rm M_{\rm BH}$/$\rm M_{\odot}$) (left panel) and $\rm log (\lambda_{Edd})$ (right panel) for the 2,290 type 1 AGNs with broad component in \oiii~(histogram filled with red line) and 3,611 type 1 AGNs without broad component (histogram filled with blue line).}
\label{comp}
\end{figure*}

As shown in Figure \ref{comp}, the two samples exhibit very similar distributions of black hole mass and Eddington ratio.
Here, the virial black hole mass and Eddington ratio of the two sample are calculated using the same method.
The sample of type 1 AGN with outflows spans $\log(M_{\rm BH}/M_{\odot})$ 
spans $6.03$-$9.34$ with a mean of 7.52 (standard deviation: 0.50), while the control sample covers $5.85$-$9.35$ with a mean of 7.52 (standard deviation: 0.55).
The values of Eddington ratios $\log (\lambda_{\rm Edd})$ range from –3.27 to 0.20 for AGNs with outflows and from –3.39 to 0.26 for the control sample without outflows. The mean values are also comparable, at –1.08 for the outflow sample and –1.00 for the control sample, with nearly identical standard deviations of 0.53 and 0.52, respectively.
The broadly similar distributions of black hole mass and Eddington ratio between the two samples indicate that the presence of [O~{\sc iii}] outflows does not substantially alter the global radiative accretion properties. Such behavior is consistent with theoretical studies showing that the kinetic power of AGN-driven winds typically constitutes only a small fraction of the bolometric luminosity \citep{Ho10,Ha18,Is18}. In this scene, outflows represent an energetically subdominant channel for dissipating accretion energy, primarily carrying it away in kinetic form without appreciably reducing the observed luminosity.

In future study, it will be important to explore the differences in the properties of the \oiii~broad component between ordinary type 1 AGNs and type 1 AGNs that exhibit double-peaked narrow emission lines. For the former, broad components are generally interpreted as signatures of AGN-driven outflows originating in the NLR, while in the latter, the observed profiles may additionally reflect complex kinematics such as dual AGN systems, rotating disks, or strong outflows. A systematic comparison of the broad components kinematics between these two classes of AGNs could provide new insights into the physical origin of the broad component and the role of galaxy environment or merger processes in shaping the \oiii~ emission profile.

\section{Summary and Conclusions}
We systematically select type 1 AGNs with \oiii~broad components from a parent sample of 11,557 QSOs at $z<0.3$ in SDSS DR16. By applying careful Gaussian profile fitting, we construct a final sample of 2,290 type 1 AGNs exhibiting broad components in the \oiii$\lambda5007$ emission line. The main results and conclusions are summarized as follows.
\begin{itemize}
\item According to the BPT diagnostic diagram, $\sim$68.21\% of our sample lie in the AGN region, while 17.55\% and 14.24\% are distributed in the star-forming and composite regions, respectively.
\item Among the sample, 1,922 objects exhibit blue-shifted broad components, 290 display symmetric broad components with respect to the core centroid, and 78 show red-shifted broad components.
\item The \oiii~broad-to-core flux ratio ranges from 0.20 to 4.56, with a median of 0.72, suggesting that the core component typically remains stronger despite the presence of notable broad components.
\item For our sample, the core component has line widths ($\sigma$) of 76.07- 315.68 \kms (mean 134.77 \kms), while the broad component spans 157.54-1068.04 \kms\ (mean 380.37 \kms). This clear difference indicates that the broad components originate from distinct regions within the NLR.
\item We measure the velocity difference $\Delta\upsilon$ between the \oiii~core and broad components, finding a distribution with a mean of –180.86 \kms, and a range from -976.13 to 309.53 \kms. A weak anti-correlation is observed between $\Delta\upsilon$ and the line width $\sigma$ (Pearson coefficient = –0.33).
\item If adopting the line width $\sigma$ of the \oiii~core component as a proxy for $\sigma_{\ast}$, the estimated black hole masses through the $M_{\rm BH}-\sigma_{\ast}$ relation shows a strong correlation with the blue emission parameter, characterized by a Pearson coefficient of –0.65. A similar correlation has also been reported in previous studies \citep{Sc18,Sc21}.
\item When virial black hole masses are used, the correlation between $M_{\rm BH}$ and the blue emission parameter becomes negligible.
\item The velocity shifts ($\Delta\upsilon$) of both the broad and core components of the \oiii~emission line show a weak correlation with the Eddington ratio, with a Pearson coefficient of –0.35 and a p-value below $10^{-10}$. This trend is consistent with the expectation that higher accretion rates enhance radiative pressure from the accretion disk, which in turn can drive faster or more pronounced outflows. 
\item If an outflow possesses a preferred axis—as in the case of jet-driven flows—variations in the line-of-sight angle can cause intrinsically similar outflows to exhibit a wide range of observed radial velocities. Consequently, even after accounting for the possible influence of radio jets, the correlation between $\Delta\upsilon$ and the Eddington ratio remains essentially unchanged from that obtained for the full sample of type 1 AGNs.
\item The similar black hole mass and Eddington-ratio distributions for type 1 AGNs with and without outflows imply that [O~{\sc iii}] outflows do not significantly affect the global radiative accretion properties. The kinetic power of AGN outflows typically accounts for only a small fraction of the bolometric luminosity, suggesting that outflows constitute an energetically subdominant feedback channel.
\item Our analysis provides new insight into how \oiii~kinematics trace the interplay between AGN accretion processes and ionized gas dynamics in the NLR.
In future work, a detailed comparison between the \oiii~broad components of typical type 1 AGNs and those of type 1 AGNs exhibiting double-peaked narrow emission lines will be an important next step. Such an analysis may uncover systematic differences in kinematic structure and asymmetry, thereby advancing our understanding of the physical mechanisms that drive outflows and the potential influence of dual AGNs on the dynamics of the NLR.
\end{itemize}

\section*{Acknowledgements}
We gratefully acknowledge the anonymous referee for giving us constructive comments and suggestions to greatly improve our paper.
This work is supported by the National Natural Science Foundation of China (Nos. 12273013, 12173020, 12373014). We have made use of the data from SDSS DR16.
The SDSS DR16 web site is (http://skyserver.sdss.org/dr16/en/home.aspx), and the SQL Search tool can be found in (\url{http://skyserver.sdss.org/dr16/en/tools/search/sql.aspx}).
The MPFIT website is (\url{http://cow.physics.wisc.edu/~craigm/idl/idl.html}).


\end{document}